\pdfoutput=1
\newcommand*{\ATLASLATEXPATH}{latex/}
\documentclass[UKenglish,texlive=2011,txfonts,cernpreprint]{\ATLASLATEXPATH atlasdoc}
\usepackage{\ATLASLATEXPATH atlaspackage}
\usepackage{\ATLASLATEXPATH atlasbiblatex}

\usepackage{\ATLASLATEXPATH atlascontribute}

\usepackage{\ATLASLATEXPATH atlasphysics}
\usepackage{\ATLASLATEXPATH atlasunit}
\graphicspath{{logos/}{figures/}}

\usepackage{fp} %used for rounding in the defs.sty
\usepackage{dcolumn}
\newcolumntype{d}[1]{D{.}{\cdot}{#1} }

\usepackage{PolSpinPaper-defs}
\usepackage{rotating}
\usepackage{multirow}

\AtlasTitle{Measurements of top quark spin observables in $t\bar{t}$ events using dilepton final states in $\sqrt{s} = 8$ \TeV\ $pp$ collisions with the ATLAS detector}

\author{The ATLAS Collaboration}

\AtlasRefCode{TOPQ-2015-13}

\PreprintIdNumber{CERN-EP-2016-263}

\AtlasJournalRef{JHEP 03 (2017) 113}
\AtlasDOI{doi:10.1007/JHEP03(2017)113}

\AtlasAbstract{%
        Measurements of top quark spin observables in
        $t \bar{t}$ events are presented based on 20.2 \invfb of $\sqrt{s} = 8$ \TeV\ proton--proton collisions recorded with the ATLAS
        detector at the LHC.
        The analysis is performed in the dilepton final state, characterised
        by the presence of two isolated leptons (electrons or muons).
        There are 15 observables,
        each sensitive to a different coefficient of the spin density
        matrix of $t\bar{t}$ production, which are measured independently. Ten of these observables are measured for the first time.
        All of them
        are corrected for detector resolution and acceptance effects
        back to the parton and stable-particle levels.
        The measured values of the observables at parton level are compared to Standard Model
        predictions at next-to-leading order in QCD.
        The corrected distributions at stable-particle level
        are presented and the means of the distributions are compared
        to Monte Carlo predictions.
        No significant deviation from the Standard Model is observed for
       any observable.
}

\hypersetup{pdftitle={ATLAS draft},pdfauthor={The ATLAS Collaboration}}

\begin{document}

\maketitle

\tableofcontents

%-------------------------------------------------------------------------------
\section{Introduction}
\label{sec:intro}
%-------------------------------------------------------------------------------

The top quark, discovered in 1995 by the CDF and D0 experiments at the
Tevatron at Fermilab~\cite{topdisc1,topdisc2}, is the heaviest fundamental
particle observed so far. 
Its mass is of the order of the electroweak scale, which suggests that it might play a special role in
electroweak symmetry breaking. Furthermore, since the top quark has a very short lifetime of
$\mathcal{O}(10^{-25}\text{
  s})$~\cite{lifetime_theo,lifetime_d0,lifetime_cdf} it decays before
hadronisation and before any consequent spin-flip can take place. This offers
a unique opportunity to study the properties of a bare
quark and, in particular, the properties of its spin.

Top quarks at the LHC are mostly produced in \ttbar pairs via the strong
interaction,
which conserves parity. The quarks\footnote{Antiparticles are generally included in the discussions unless otherwise stated.} and gluons of the initial state
are unpolarised, which means that their spins are not preferentially aligned with any given direction.
The top quarks produced in pairs are thus unpolarised except for the contribution of weak corrections and QCD absorptive
parts at the per-mill level~\cite{bernreuther_spin2013}. However, the spins of the top and antitop
quarks are correlated with a strength depending on the spin
quantisation axis and on the production process. Various new physics
phenomena can alter the polarisation and spin correlation due to
alternative production
mechanisms~\cite{bernreuther_spin2013,spin_obs_theo,godbole_chargedhiggs,godbole_sbottom}.
The spins of the top quarks do not become decorrelated
due to hadronisation, and so their spin information is transferred to their decay
products. This makes it possible to measure the top quark pair's 
spin structure using angular observables of their decay products.
The predictions for many of these observables are available at next-to-leading order (NLO) in quantum chromodynamics (QCD). 
A few of them have been measured by the experiments at the LHC and Tevatron and found to be in good agreement with the Standard Model (SM)
predictions~\cite{spin_atlas1,pol_atlas,Aad:2015bfa,spin_atlas2,spinpol_cms,Khachatryan:2015tzo,spinpol_cms8TeV,pol_d0_ljets,spin_cdf}.

This paper presents the measurement of a set of 15 spin observables with
a data set corresponding to an integrated luminosity of $20.2\text{ fb}^{-1}$ of proton--proton collisions
at $\sqrt{s} = 8$ \TeV, recorded by the ATLAS detector at the LHC in 2012. Each
of the 15 observables is sensitive to a different coefficient of the top
quark pair's spin density matrix, probing different symmetries in the
production mechanism~\cite{bernreuther_observables}. Ten of these observables have not
been measured until now. The observables are corrected back to parton level in the full phase-space and
to stable-particle level in a fiducial phase-space.
At parton level, the measured values of the polarisation
and spin correlation observables are presented and compared to
theoretical predictions. All observables allow a direct
measurement of their corresponding expectation value.
At stable-particle level, the distributions corrected for detector
acceptance and resolution are provided. Because of the limited phase-space used at that level,
the values of the polarisation and spin correlations
are not proportional to the means of these distributions.
Instead, the means of the distributions are provided and compared to the values obtained in Monte Carlo simulation.

\section{ATLAS detector}
\label{sec:detector}
The ATLAS detector~\cite{PERF-2007-01} at the LHC covers nearly the entire solid angle around the interaction
point.\footnote{ATLAS uses a right-handed coordinate system with its origin at the nominal interaction point (IP) in
the centre of the detector and the $z$-axis along the beam pipe. 
The $x$-axis points from the IP to the centre of the LHC ring, and the $y$-axis points upward. Cylindrical coordinates $(r, \phi)$ are used in the transverse
plane, $\phi$ being the azimuthal angle around the beam pipe. The pseudorapidity is defined in terms of 
the polar angle $\theta$ as $\eta = - \ln[\tan (\theta/2)]$.}
It consists of an inner tracking detector surrounded by a thin superconducting solenoid, electromagnetic and hadronic calorimeters, and a muon spectrometer incorporating
 superconducting toroid magnets. 

The inner-detector system is immersed in
a $2$~T axial magnetic field and provides charged-particle tracking in
the pseudorapidity 
range $|\eta| <2.5 $. 
A high-granularity silicon pixel detector covers the interaction region and typically
provided three measurements per track in 2012. It is surrounded by a silicon microstrip tracker
designed to provide eight two-dimensional measurement points per track. These silicon detectors
are complemented by a transition radiation tracker, which enables radially extended
track reconstruction up to $|\eta| =2.0 $. The transition radiation tracker also provides electron
identification information based on the fraction of hits (typically 30 in total) exceeding an
energy-deposit threshold consistent with transition radiation.

The calorimeter system covers the pseudorapidity range $|\eta| <4.9 $. Within the region
$|\eta| <3.2 $, electromagnetic calorimetry is provided by barrel and endcap high-granularity
lead/liquid-argon (LAr) electromagnetic calorimeters, with an additional thin LAr presampler
covering $|\eta| <1.8 $ to correct for energy loss in the material upstream of the calorimeters.
Hadronic calorimetry is provided by a steel/scintillator-tile calorimeter, segmented into
three barrel structures within $|\eta| <1.7 $, and two copper/LAr hadronic endcap calorimeters.
The solid angle coverage is completed with forward copper/LAr and tungsten/LAr
calorimeters used for electromagnetic and hadronic measurements, respectively.

The muon spectrometer comprises separate trigger and high-precision tracking chambers
measuring the deflection of muons in a magnetic field generated by superconducting
air-core toroids. The precision chamber system covers the region $|\eta| <2.7 $ with drift tube chambers, 
complemented by cathode strip chambers. The muon trigger system covers 
the range $|\eta| <1.05$ with resistive plate chambers,
and the range $1.05< |\eta| <2.4$ with thin gap chambers.

A three-level trigger system is used to select interesting events. The Level-1 trigger is
implemented in hardware and uses a subset of detector information to reduce the event rate
to a design value of at most 75~kHz. This is followed by two software-based trigger levels,
which together reduce the event rate to about 400 Hz.

\section{Observables}
\label{sec:observables}

The spin information of the top quarks, encoded in the spin density matrix, is transferred to their decay particles and affects their
angular distributions. 
The spin density matrix can be expressed by a set of several coefficients:
one spin-independent coefficient, which determines the cross section and which is not measured here, three
polarisation coefficients for the top quark, three polarisation coefficients for the antitop quark, 
and nine spin correlation coefficients. 
By measuring a set of 15 polarisation and spin correlation observables, 
the coefficient functions of the squared matrix element can be probed. 
The approach used in this paper was proposed in Ref.~\cite{bernreuther_observables}.
The normalised double-differential cross section for \ttbar production
and decay is of the form~\cite{bernreuther_spin2013,Bernreuther:2004jv}
\begin{equation}
\frac{1}{\sigma} \frac{\textrm{d}^{2} \sigma}{\textrm{d} \cos \theta^{a}_{+} \textrm{d} \cos
  \theta^{b}_{-}} = \frac{1}{4} (1 +B^{a}_{+} \cos \theta^{a}_{+} +
B^{b}_{-} \cos \theta^{b}_{-} -  C(a,b) \cos
\theta^{a}_{+} \cos \theta^{b}_{-}),
\label{eqn:double-diff}
\end{equation}
where $B^{a}$, $B^{b}$ and $C(a,b)$ are the polarisations and spin correlation along the
spin quantisation axes $a$ and $b$. The angles $\theta^{a}$ and $\theta^{b}$ are defined as the angles between the momentum
direction of a top quark decay particle in its parent top quark's rest frame and the axis $a$ or $b$. The
subscript $+ (-)$ refers to the top (antitop) quark. 
From Equation~(\ref{eqn:double-diff}) one can retrieve the
following relation for the spin correlation between the axes $a$ and $b$
\begin{equation}
 C(a,b) = -9<\cos \theta_{+}^{a} \cos\theta_{-}^{b}>.
 \label{eqn:spin_relation}
\end{equation} 
Integrating out one of the
angles in Equation~(\ref{eqn:double-diff}) gives the single-differential
cross section 
\begin{equation}
  \frac{1}{\sigma} \frac{\textrm{d} \sigma}{\textrm{d} \costheta^a} = \frac{1}{2} (1 +
  B^a \costheta^a).
  \label{eqn:single-diff}
\end{equation}
This means the differential cross section has a linear dependence on the polarisation
$B^a$, from which also follows 
\begin{equation}
B^a = 3<\cos \theta^a>.
\label{eqn:pol_relation}
\end{equation}

All the observables are
based on \costheta, which is defined using three
orthogonal spin quantisation axes:
\begin{itemize}
  \item The helicity axis is defined as the top quark direction in the \ttbar rest
    frame. In Ref.~\cite{bernreuther_observables} it is
    indicated by the letter $k$, a notation which is adopted in this paper. 
    Measurements of the polarisation and spin correlation defined along
    this axis at 7 and 8 \TeV\ were consistent with the SM predictions~\cite{pol_atlas,spin_atlas1,Aad:2015bfa,spin_atlas2,spinpol_cms,Khachatryan:2015tzo, spinpol_cms8TeV}.
  \item The transverse axis is defined to be transverse to the production plane~\cite{bernreuther_spin2013,baumgart_transpol}
    created by the top quark direction and the beam axis. It is denoted by the letter $n$. 
    The polarisation along that axis was measured by the D0 experiment~\cite{pol_d0_ljets}.
  \item The $r$-axis is an axis orthogonal to the other two axes, denoted
    by the letter $r$. No observable related to this axis has been
    measured previously.
\end{itemize}

As the dominant initial state of \ttbar production at the LHC (gluon--gluon fusion) is
Bose-symmetric, \costheta calculations with respect to the transverse or $r$-axis
are multiplied by the sign of the scattering angle $y =
\hat{\mathbf{p}}\cdot \hat{\mathbf{k}}$, where $\hat{\mathbf{k}}$ is
the top quark direction in the \ttbar rest frame and $\hat{\mathbf{p}} =
(0,0,1)$, as recommended in Ref.~\cite{bernreuther_observables}. 
In the calculations of \costheta with respect to the negatively charged
lepton, the axes are multiplied by $-1$.
The observables and corresponding expectation values, as well as their SM predictions at NLO, 
are shown in Table~\ref{tab:observables}. 
The first six observables correspond to the polarisations of 
the top and antitop quarks along the various axes, the other nine
to the spin correlations. In order to distinguish between the
correlation observables, the correlations using only one
axis are referred to as spin correlations and the last six as
cross correlations.
The predictions are computed for a top quark mass of 173.34~\GeV~\cite{ATLAS:2014wva}.
In order to
measure all observables, the final-state particles of both decay
chains must be reconstructed and correctly identified. 
As charged leptons retain more information about the spin state of the top quarks, 
and as they can be precisely reconstructed, the measurement in this paper is performed in the
dileptonic final state of \ttbar events.
The charged leptons considered in this analysis are electrons or muons, either originating directly from $W$ and $Z$ decays, 
or through an intermediate $\tau$ decay.

\begin{table}
  \begin{center}
  \renewcommand{\arraystretch}{1.1}
    \begin{tabular}{ccc}
      \hline 
      Expectation values & NLO predictions & Observables \\
      \hline
      \helpolplus    &$0.0030 \pm 0.0010$ & \helcosplus         \\[0.05cm]
      \helpolminus   &$0.0034 \pm 0.0010$ & \helcosminus        \\[0.05cm]
      \transpolplus  &$0.0035\pm0.0004$   & \transcosplus       \\[0.05cm]
      \transpolminus &$0.0035\pm0.0004$   & \transcosminus      \\[0.05cm]
      \rpolplus      &$0.0013 \pm 0.0010$ & \rcosplus           \\[0.05cm]
      \rpolminus     &$0.0015 \pm0.0010$  & \rcosminus          \\[0.05cm]
      \helcorr       &$0.318 \pm 0.003$   & \helcoscos          \\[0.05cm]
      \transcorr     &$0.332 \pm 0.002$   & \transcoscos        \\[0.05cm]
      \rcorr         &$0.055 \pm 0.009$   & \rcoscos            \\[0.05cm]
      \transhelsum   &$0.0023$            & \transhelcoscossum  \\[0.05cm]
      \transheldiff  &$0$                 & \transhelcoscosdiff \\[0.05cm]
      \transrsum     &$0.0010 $           & \transrcoscossum    \\[0.05cm]
      \transrdiff    &$0$                 & \transrcoscosdiff   \\[0.05cm]
      \rhelsum       &$-0.226 \pm 0.004$  & \rhelcoscossum      \\[0.05cm]
      \rheldiff      &$0$                 & \rhelcoscosdiff     \\[0.05cm]
      \hline
      \end{tabular}
      \caption{List of the observables and corresponding expectation values measured in this analysis. 
        The SM predictions at NLO are also
        shown~\cite{bernreuther_observables}; expectation values
        predicted to be 0 at NLO are exactly 0 due to term cancellations.
        The expectation values can be obtained from the corresponding observables using the relations
        from Equations~(\ref{eqn:spin_relation}) and (\ref{eqn:pol_relation}).
        The uncertainties on the predictions refer to scale uncertainties only; values below 10$^{-4}$ are not quoted.}
   \label{tab:observables}
    \end{center}
\end{table}

\section{Data and simulation samples}
\label{sec:mcsamples}
The analysis is performed using the full 2012 proton--proton collision data sample at $\sqrt{s} = 8$ \TeV\ 
recorded by the ATLAS detector.
The data sample corresponds to an integrated luminosity of 20.2 \invfb after requiring stable LHC beams and 
a fully operational detector.

The analysis uses Monte Carlo (MC) simulations, in particular to estimate the sample composition
and to correct the measurement to both parton and stable-particle level.
The nominal \ttbar signal MC sample is generated by
\powheg (version 1, r2330)~\cite{powhegbox,powhegbox2,powhegbox3,Frixione:2007nw} with the top quark mass set to 172.5 \GeV\ and the
 \hdamp parameter\footnote{The \hdamp parameter controls the hardness of the hardest emission which recoils
against the \ttbar system~\cite{ATL-PHYS-PUB-2015-002,ATL-PHYS-PUB-2015-011}.} set to the top quark mass. The PDF set used is CT10~\cite{ct10}.
The signal events are then
showered with \pythia (version 6.426)~\cite{pythia6} using a set of tuned parameters named the Perugia2011C tune~\cite{perugia_tune}. 
The background processes are also modelled using a range of MC generators which are listed in
Table~\ref{tab:mc_gens_bkg}. An additional background originating from
non-prompt and misreconstructed (called ``fake'') leptons is also estimated from MC simulation.
To estimate this background, all samples listed in Table~\ref{tab:mc_gens_bkg} are used, and in particular those listed in the lower part of the table, 
which are generated specifically for that background. Multijet events are not included in this list 
because the probability of having two jets misidentified as isolated leptons is very small. 
The contribution from these events is thus negligible.

In order to account for systematic uncertainties in
the signal modelling, different MC samples, documented in Table~\ref{tab:mc_gens_signal}, 
are compared with each other as described in Section~\ref{sec:modellingsysts}.

The nominal signal and background samples were processed through a simulation
of the detector geometry and response~\cite{Aad:2010ah} using \geant~\cite{geant4}.
MC samples used to estimate signal modelling uncertainties were processed with the \atlfast~\cite{ATLAS:1300517}
simulation. This employs a parameterisation of the response of the electromagnetic and 
hadronic calorimeters, and uses \geant for the other detector components.

\begin{table}
  \begin{center}
    \renewcommand{\arraystretch}{1.1}
    \begin{tabular}{lcccc}
      \hline
      Process & Generator & Parton shower & PDF set & Tune \\
      \hline
      \ttbar                  & \powheg~\cite{Re:2010bp} & \pythia & CT10~\cite{ct10} & Perugia2011C~\cite{perugia_tune} \\
      Single top ($Wt$-channel)& \powheg & \pythia & CT10 & Perugia2011C \\
      Drell--Yan               & \alpgen & \pythia & CTEQ6L1~\cite{Pumplin:2002vw} & Perugia2011C \\
      Dibosons ($WW$,$WZ$,$ZZ$)& \alpgen & \herwig\!6+\jimmy & CTEQ6L1 & AUET2~\cite{ATL-PHYS-PUB-2011-008} \\
      $t\bar{t}V$ ($V = W/Z/\gamma^*$)& \madgraph & \pythiaeight & CTEQ6L1 & AUET2B \\
      \hline
      Single top, t-channel    & \acermc & \pythia & CTEQ6L1 & Perugia2011C \\
      $W$+jets                 & \alpgen & \pythia & CTEQ6L1 & Perugia2011C \\
      $W+\gamma$+jets          & \alpgen & \herwig\!6+\jimmy & CTEQ6L1 & AUET2  \\
      \hline
    \end{tabular}
    \caption{MC generators and parton showers used for the signal and background processes. 
      Samples in the lower part of the table are used together with the other samples to estimate the fake-lepton background.
      The parton distribution functions (PDF) used by the generator and the tunes used for the parton shower are also shown.
      The versions of the different generators are 2.14 for \alpgen~\cite{alpgen}, 5.1.4.8 for \madgraph~\cite{madgraph}, 
      4.31 for \herwig\!6+\jimmy~\cite{jimmy}, 1, r2330 for \powheg, 6.426 for \pythia~\cite{pythia6} and 8.165 for \pythiaeight~\cite{pythia8}.}
    \label{tab:mc_gens_bkg}
  \end{center}
\end{table}

\begin{table}
  \begin{center}
    \renewcommand{\arraystretch}{1.1}
    \begin{tabular}{lccc}
    \hline
     Systematic uncertainty & Generator & Parton shower & Tune   \\
      \hline
     \multirow{2}{*}{Colour reconnection} & \multirow{2}{*}{\powheg} & \multirow{2}{*}{\pythia} & Perugia2012~\cite{perugia_tune}   \\
                                          &                          &                          & Perugia2012loCR   \\ \hline
     \multirow{2}{*}{Underlying event}    & \multirow{2}{*}{\powheg} & \multirow{2}{*}{\pythia} & Perugia2012   \\
                                          &                          &                          & Perugia2012mpiHi \\ \hline
     \multirow{2}{*}{Parton shower}       & \multirow{2}{*}{\powheg} & \pythia                  & Perugia2011C \\
                                          &                          & \herwig                  & AUET2 \\ \hline
     \multirow{2}{*}{Generator}           & \powheg                  & \multirow{2}{*}{\herwig} & \multirow{2}{*}{AUET2} \\
                                          & \mcatnlo                 &                          &                         \\ \hline
     \multirow{2}{*}{ISR/FSR}             & \multirow{2}{*}{\powheg} & \multirow{2}{*}{\pythia} & Perugia2012radLo \\ 
                                          &                          &                          & Perugia2012radHi \\ \hline
     Top-quark mass                       & \powheg                  & \pythia                  & Perugia2011C with various mass points \\                                     
     \hline
    \end{tabular}
    \caption{List of \ttbar samples used for studies of the modelling uncertainties. 
        The PDF set is CT10 for all of them.
        The version of the \mcatnlo generator~\cite{mcatnlo1,mcatnlo2} is 4.01. 
    }
    \label{tab:mc_gens_signal}
  \end{center}
\end{table}

\section{Event selection and background estimation}
\label{sec:selection}

Reconstructed objects such as electrons, muons or jets
are built 
from the detector information and used to form a \ttbar-enriched
sample by applying an event selection.

\subsection{Object selection}
Electron candidates are reconstructed by matching inner-detector tracks to clusters in
the electromagnetic calorimeter. A requirement on the pseudorapidity of the
cluster $|\eta_{\text{cl}}| < 2.47$ is applied, with the
transition region between barrel and endcap
corresponding to
$1.37 < |\eta_{\text{cl}}| < 1.52$ excluded. A minimum requirement on the transverse
momentum (\pt) of 25~\GeV\ is applied to match the trigger criteria (see Section~\ref{sec:obj_sel}). 
Furthermore, electron candidates are required to be isolated from additional activity in the
detector. Two different criteria are used. The first one considers the activity in the electromagnetic calorimeter in a cone of size $\Delta R = 0.2$ around the electron.
The second one sums the $p_T$ of all tracks in a cone of size 0.3 around the electron track. 
The requirements applied on both variables are $\eta$-dependent and correspond to an efficiency on signal electrons of 90$\%$. 
The final selection efficiency for the electrons used in this
analysis is between 85\% and 90\% depending on the \pt and $\eta$ of the
electron~\cite{elec_eff}. 
%The probability for a fake electron to be accepted with this selection is around 0.5\%.

Muon candidates are reconstructed by combining inner detector tracks with
tracks constructed in the muon spectrometer. They are required to
have a \pt $> 25$ \GeV\ and $|\eta| < 2.5$. 
They are also required to be isolated from additional activity in the inner detector.
An isolation criterion
requiring the scalar sum of track \pt around the muon in a cone of size 
$\Delta R = 10 \text{ \GeV}/p_{\text{T}}^{\mu}$ to be less than
$0.05p_{\text{T}}^{\mu}$ is applied. 
Muons have a selection 
efficiency of about 95\%~\cite{Aad:2014rra}.

Jets are reconstructed from energy clusters in the electromagnetic and
hadronic calorimeters. The reconstruction algorithm used is the
anti-$k_{t}$~\cite{antikt_ref} algorithm with a radius parameter of
$R=0.4$. 
The measured energy of the jets is corrected to the hadronic scale using \pt- and $\eta$-dependent scale factors derived from simulation and validated in
data~\cite{atlas_jes}.
After the energy correction, they are required to have a transverse momentum $p_{\text{T}}
> 25$ \GeV\ and a pseudorapidity $|\eta| < 2.5$. For jets with
$p_{\text{T}} < 50$ \GeV\ and $|\eta| < 2.4$, the jet vertex fraction
(JVF) must be greater than 0.5. The JVF is defined as the fraction
of the scalar \pt sum of tracks associated with the jet and the primary vertex 
and the scalar \pt sum of tracks associated with the jet and any vertex. It 
distinguishes between jets originating from the primary vertex and jets
with a large contribution from other proton interactions in the same bunch crossing (pile-up).
If separated by $\Delta R < 0.2$, the jet closest to a
selected electron is removed to avoid double-counting of electrons reconstructed as jets. 
Next, all electrons and muons separated from a jet by $\Delta R < 0.4$ are removed 
from the list of selected leptons to reject semileptonic decays within a jet. Jets containing $b$-hadrons are identified
($b$-tagged) by using a multivariate algorithm
(MV1)~\cite{Aad:2015ydr} which uses
information about the tracks and secondary vertices. If the MV1 output for a
jet is larger than a predefined value, the jet is considered to be 
$b$-tagged. The value was chosen to achieve a $b$-tagging
efficiency of 70\%. With this algorithm, the probability to select 
a light jet (from gluons or $u$-, $d$-, $s$-quarks) is around 0.8\%, and the probability
to select a jet from a $c$-quark is 20\%.%, evaluated from a \ttbar MC sample.

The missing transverse momentum \met is defined
as the magnitude of the negative vectorial sum of the transverse momenta of leptons, photons 
and jets, as well as energy deposits in the calorimeter not  
associated with any physics object~\cite{atlas_met}.

\subsection{Event selection}
\label{sec:obj_sel}
The event selection aims at maximising the fraction of \ttbar
events with a dileptonic final state. 
The final states are then separated according to the 
lepton flavours. Tau leptons  
are indirectly considered in the signal contribution when decaying leptonically. This leads to three
different channels (\eeme, \mumume, \emume). Different
kinematic requirements have to be applied for the \emume and \eeme/\mumume
channels due to their different background contributions. Only events selected
from dedicated electron or muon triggers are considered. The \pt
thresholds of the triggers are 24 \GeV\ for isolated leptons and 60 (36)
\GeV\ for single-electron (-muon) triggers without an isolation requirement. 
Events containing muons compatible with cosmic-ray interactions are removed.
Exactly two oppositely
charged electrons or muons with  
$p_{\text{T}} > 25$ \GeV\ are required.
A requirement on the dilepton invariant mass
of $m_{\ell\ell} > 15$ \GeV\ is required in all channels.
In addition, $|m_{\ell\ell} - m_{Z}| > 10$ \GeV,
where $m_{Z}$ is the $Z$ boson mass, is
required in the \eeme and \mumume channels to suppress the Drell--Yan
background. In these channels the missing transverse
momentum is required to be greater than 30~\GeV. In the \emume channel, the scalar
sum of the $p_{\text{T}}$ of the jets and leptons in the event
($H_{\text{T}}$) is required to be $H_{\text{T}} > 130$
\GeV. At least two jets with at least one of them being $b$-tagged are required in each channel.

\subsection{Background estimation}
\label{sec:bkg}

Single-top-quark and diboson backgrounds are estimated using MC simulation only.
The MC estimate for the Drell--Yan and fake-lepton background is normalised 
using data-driven scale factors (SF).
The Drell--Yan background does not contain any real \met. 
Non-negligible \met can appear in a fraction of events with
misreconstructed objects, which are difficult to model. 
Since real \met is present in $Z \rightarrow \tau\tau$ events,
no scale factors are applied to this sample.
Another issue is the correct normalisation of
Drell--Yan events with additional heavy-flavour (HF) jets from $b$- and $c$-quarks after the
$b$-tagging requirement.
In order to correct for these effects, three 
control regions are defined, from which three SF are extracted. 
Two correspond to the \met modelling in $Z \rightarrow ee$ (SF$_{ee}$) and $Z \rightarrow \mu\mu$ events
(SF$_{\mu\mu}$), and one for the heavy-flavour normalisation in
$Z$+jets events (SF$_{\text{HF}}$)
common to the three dilepton channels. 
All control regions require the same
selection as the signal region with the exception that the invariant mass of the 
two leptons should be within 10~\GeV\ of the $Z$ mass.
The control regions are then distinguished by dividing them
into a pretag ($n_{b\mhyphen\text{tag}} \geq 0$) and a $b$-tag region
($n_{b\mhyphen\text{tag}} \geq 1$), additionally dividing the pretag
region into the \eeme and \mumume channels.
The purity of the pretag control region is 97$\%$ on average for both channels. The purity of the $b$-tag region is 75$\%$.
The SF are extracted by solving a system of equations which relates 
the number of events in data and in simulation in the three control regions.
The lepton-flavour-dependent scale factors SF$_{ee/\mu\mu}$ are $0.927 \pm 0.005$
and $0.890 \pm 0.004$ respectively for the $ee$ and $\mu\mu$ channels 
while the heavy-flavour scale factor SF$_{\text{HF}}$ is $1.70 \pm 0.03$, where the uncertainties
are only statistical.

The shape of the fake and non-prompt lepton background distributions are taken from 
MC simulation but the normalisation is derived from data in a control 
region enriched in fake leptons.
This is achieved by applying the same requirements as for the
signal region, except that two leptons of the same charge are required. As
fake leptons have approximately the same probability of having
negative or positive charge, the same number of fake-lepton events should 
populate the opposite-sign and same-sign selection regions. The
same-sign control region has a smaller background contribution from
other processes, allowing the study of the modelling of the fake-lepton background. Channel-dependent scale factors are derived
by normalising the predictions to data in the control regions, while the
shapes of the distributions are taken from MC simulation. The SF in the \eeme and \emume channels are around 1.0 and 1.5, whereas the SF in the \mumume
channel is
about 4. The differences between the three scale factors originate from the 
sources of misidentified electrons and muons, which seem to be
modelled better in MC simulation for the electrons. However, the shapes of the
distributions of several kinematic variables in the \mumume channel are cross-checked in control regions and found to be consistent with the distributions from a purely data-driven method. The relative statistical uncertainties are about 20\% in the
same-flavour channels and 10\% in the \emume channel.

\subsection{Kinematic reconstruction of the $t\bar{t}$ system}
\label{sec:reco}
The dileptonic \ttbar final state consists of two charged leptons, two
neutrinos and at least two jets originating from the top quark 
decay. As the neutrinos cannot be directly observed in the detector, the kinematics
of the \ttbar
system, which is necessary to construct the observables, cannot be
simply reconstructed from the measured information. To solve the
kinematic equations and reconstruct the \ttbar system, the neutrino weighting technique~\cite{Abbott:1997fv,Aad:2015jfa} is used.

As input, the method uses the measured lepton and jet momenta. 
The masses of the top quarks are set to their generated mass of 172.5
\GeV\ whereas the masses of the $W$ bosons 
are set to their PDG values~\cite{Olive:2016xmw} in the
calculations.
A hypothesis is made for the value of the pseudorapidity of each neutrino and 
the kinematics of the system is then solved. 
For each solution found, a weight is assigned to quantify the level of agreement between 
the vectorial sum of neutrino transverse momenta and the measured \met components.
The pseudorapidities of the neutrinos are scanned independently between \mbox{$-5$} and 5 
with fixed steps of 0.025 in the range [$-2$, 2] and of 0.05 outside of that range.
All possible combinations of jets and leptons are tested.
Additionally, the resolution of the jet energy measurement is taken into account by smearing the energy of each jet 50 times. The smearing is done using transfer functions mapping the energy at particle level
to the energy after detector simulation.
Out of all the solutions obtained, the one with the highest weight is selected.
The reconstruction efficiency of the kinematic
reconstruction in the \ttbar signal sample is about 88\%.
No solution is found for the remaining events.

\subsection{Event yields and kinematic distributions}

\begin{figure}[!htbp]
 \begin{center}
   \includegraphics[scale=0.33]{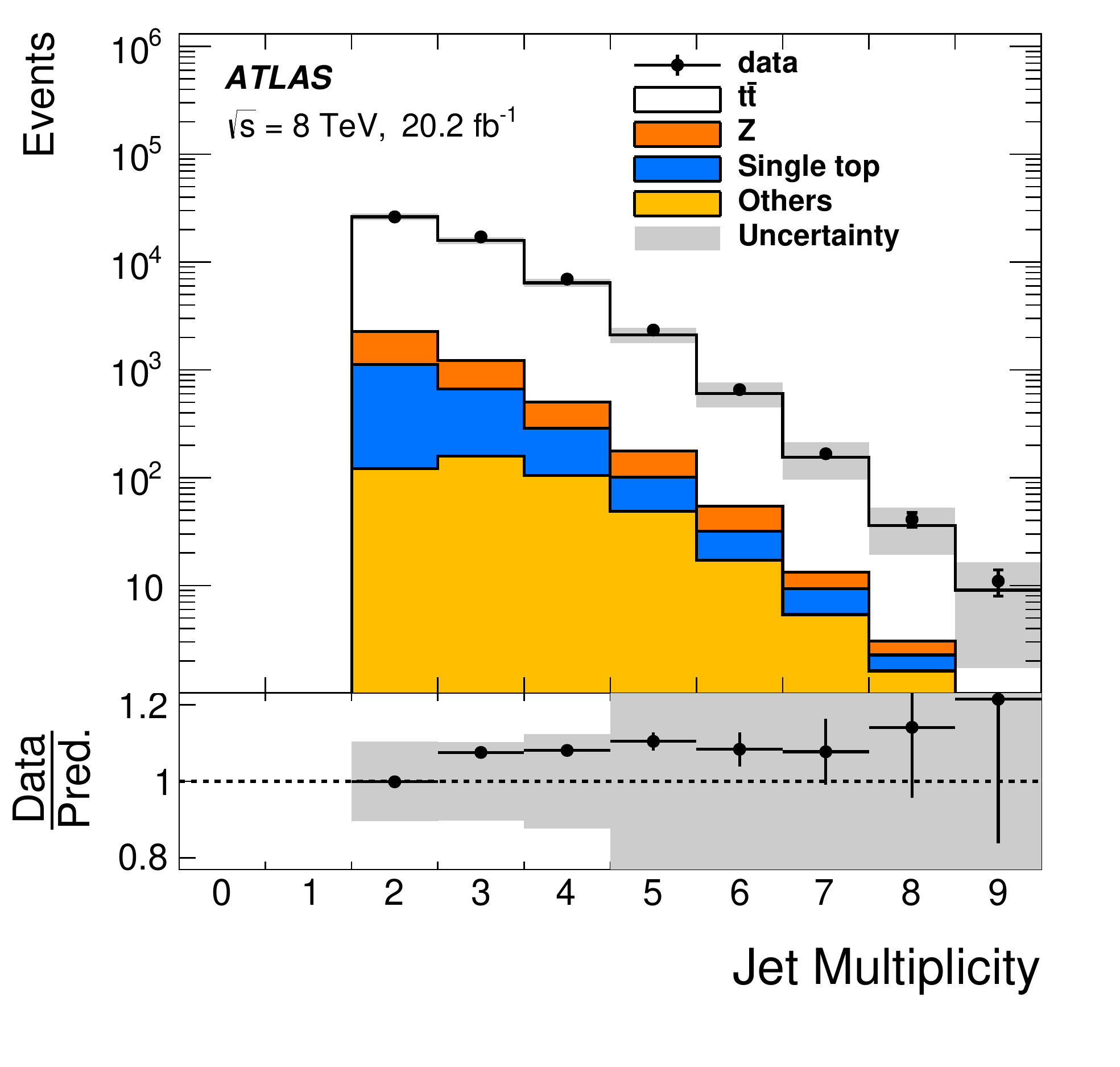}\\
   \includegraphics[scale=0.33]{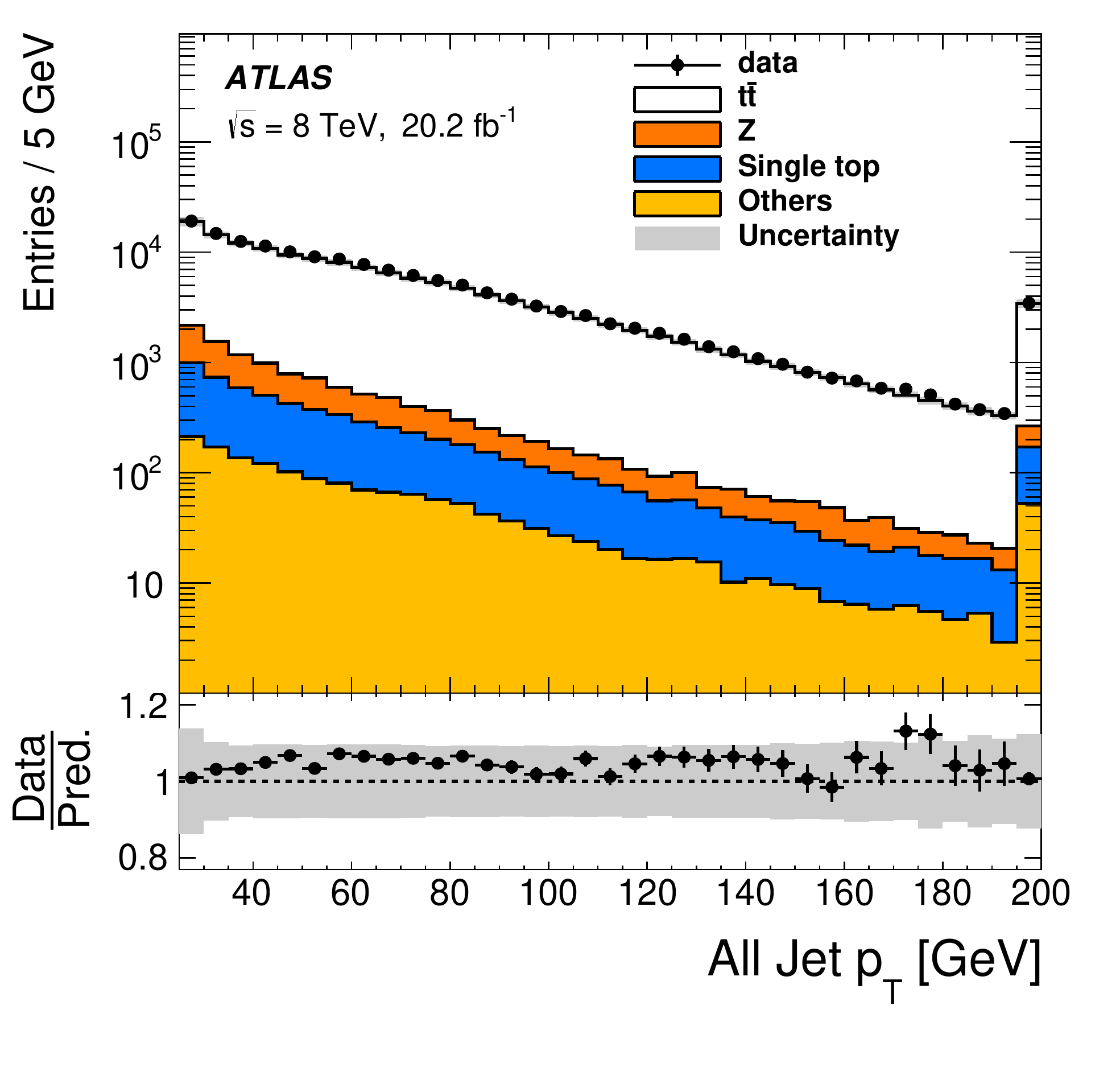}
   \includegraphics[scale=0.33]{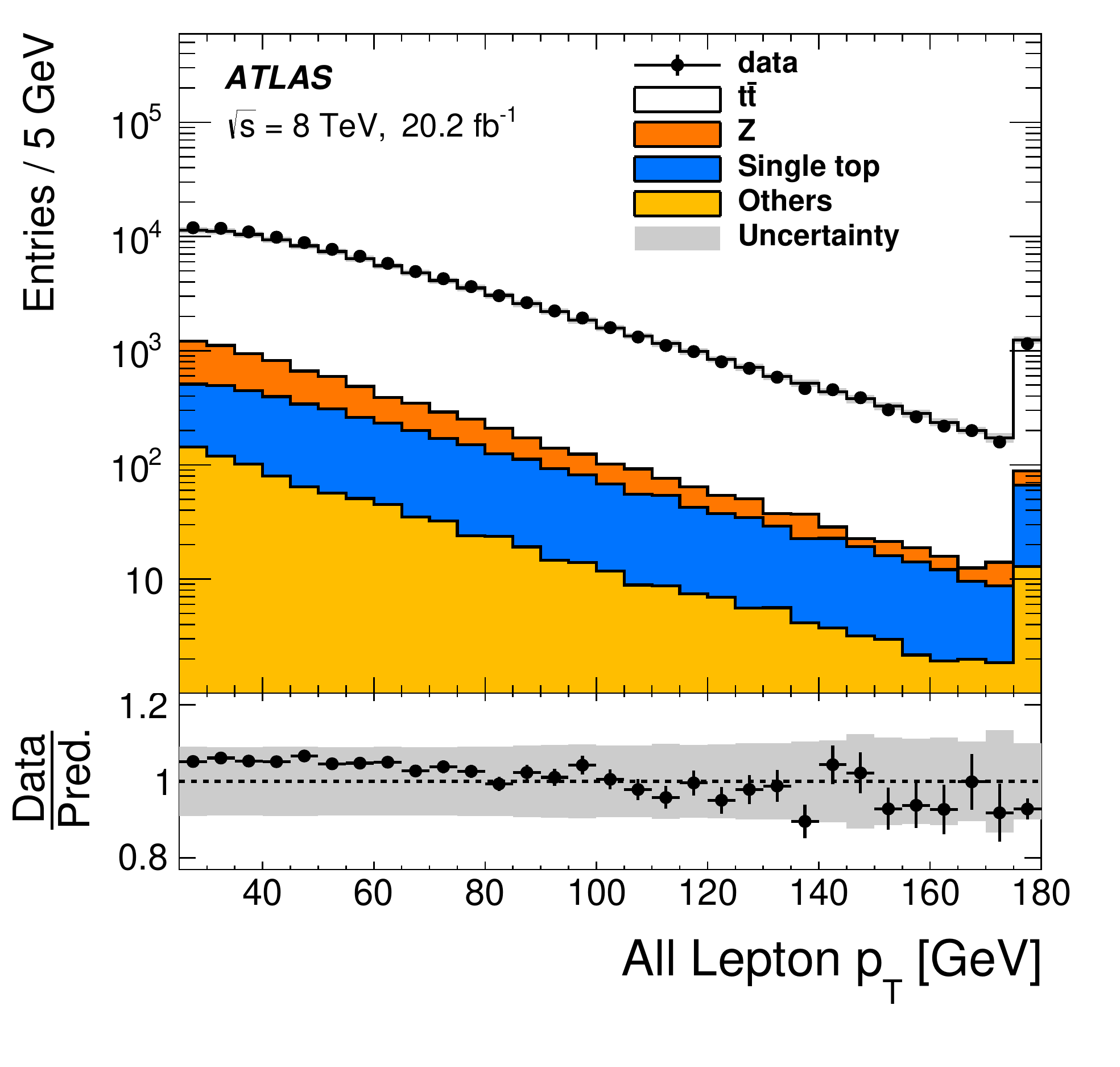}\\
   \caption{Comparison of the number of jets,
     jet \pt and lepton \pt distributions between data and
     predictions after the event
     selection in the combined dilepton channel. The ratio between the data and prediction is also shown. The grey
     area shows the statistical and systematic uncertainty on the signal and background. 
     The $t\bar{t}V$, diboson and fake-lepton backgrounds are shown together in the "Others" category.
     Only the events passing the kinematic reconstruction are considered in the distributions.} 
   \label{fig:sel_dataMC}
 \end{center}
\end{figure}

\begin{figure}[!htbp]
 \begin{center}
   \includegraphics[scale=0.33]{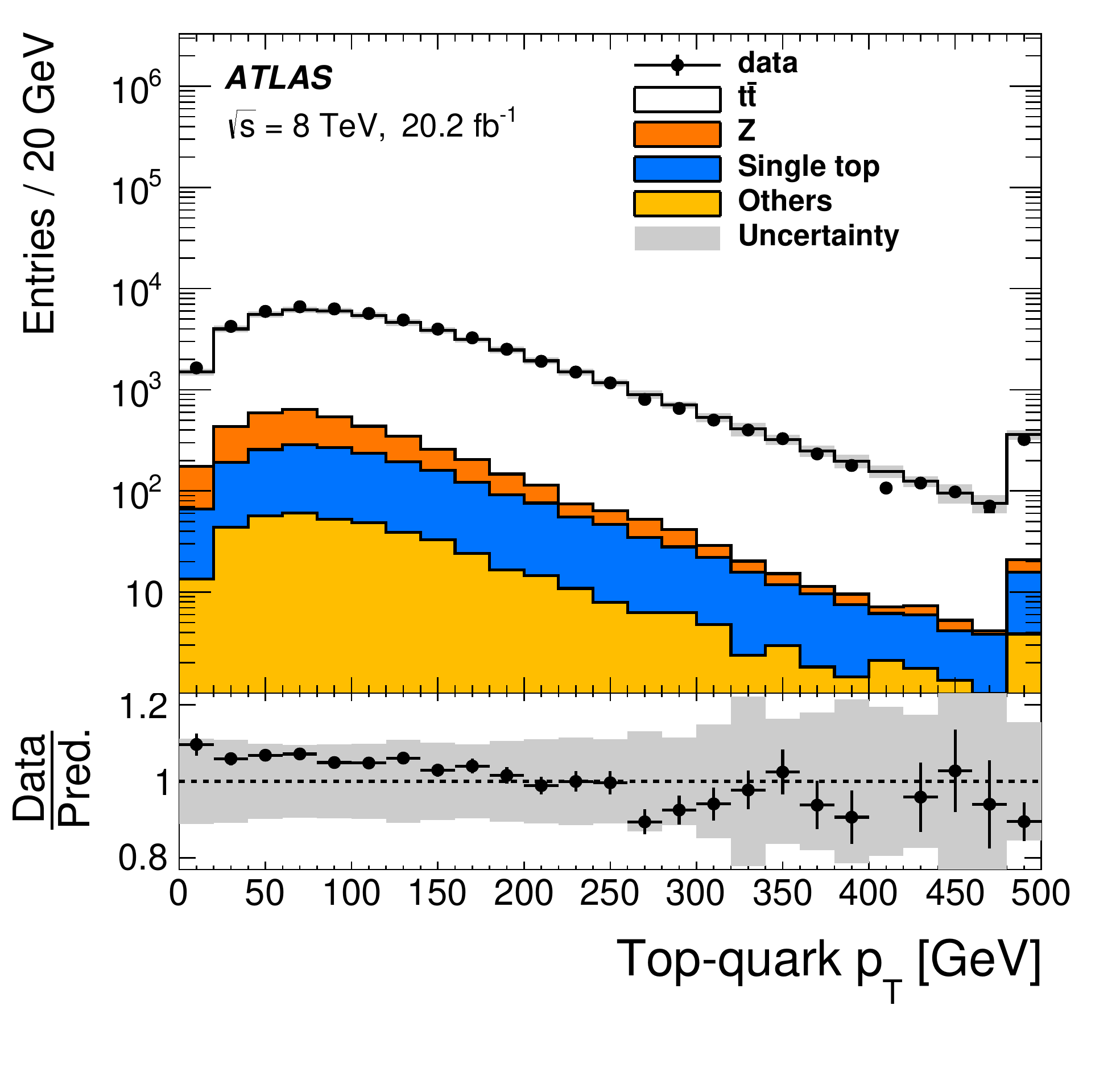}
   \includegraphics[scale=0.33]{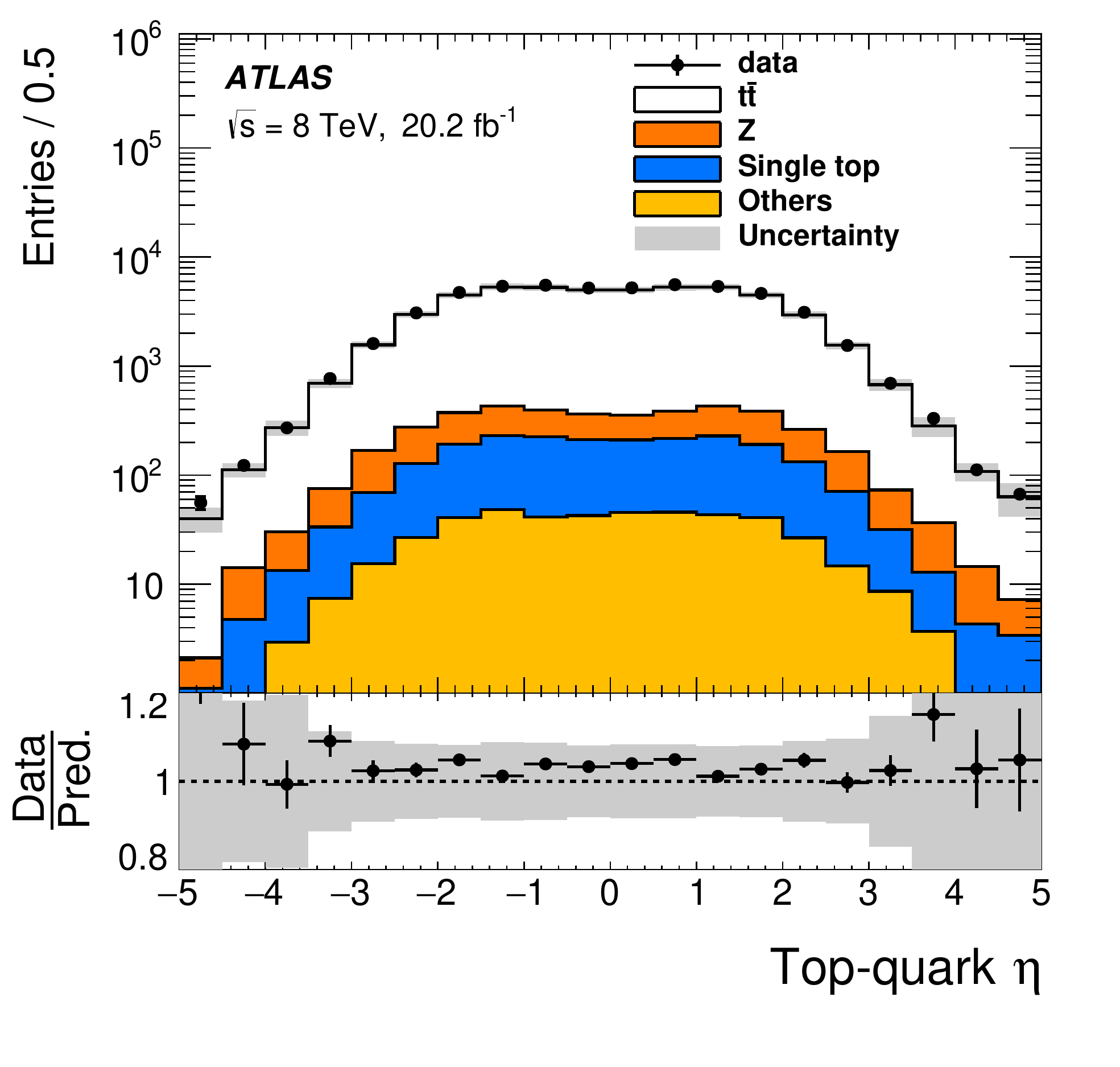}
   \includegraphics[scale=0.33]{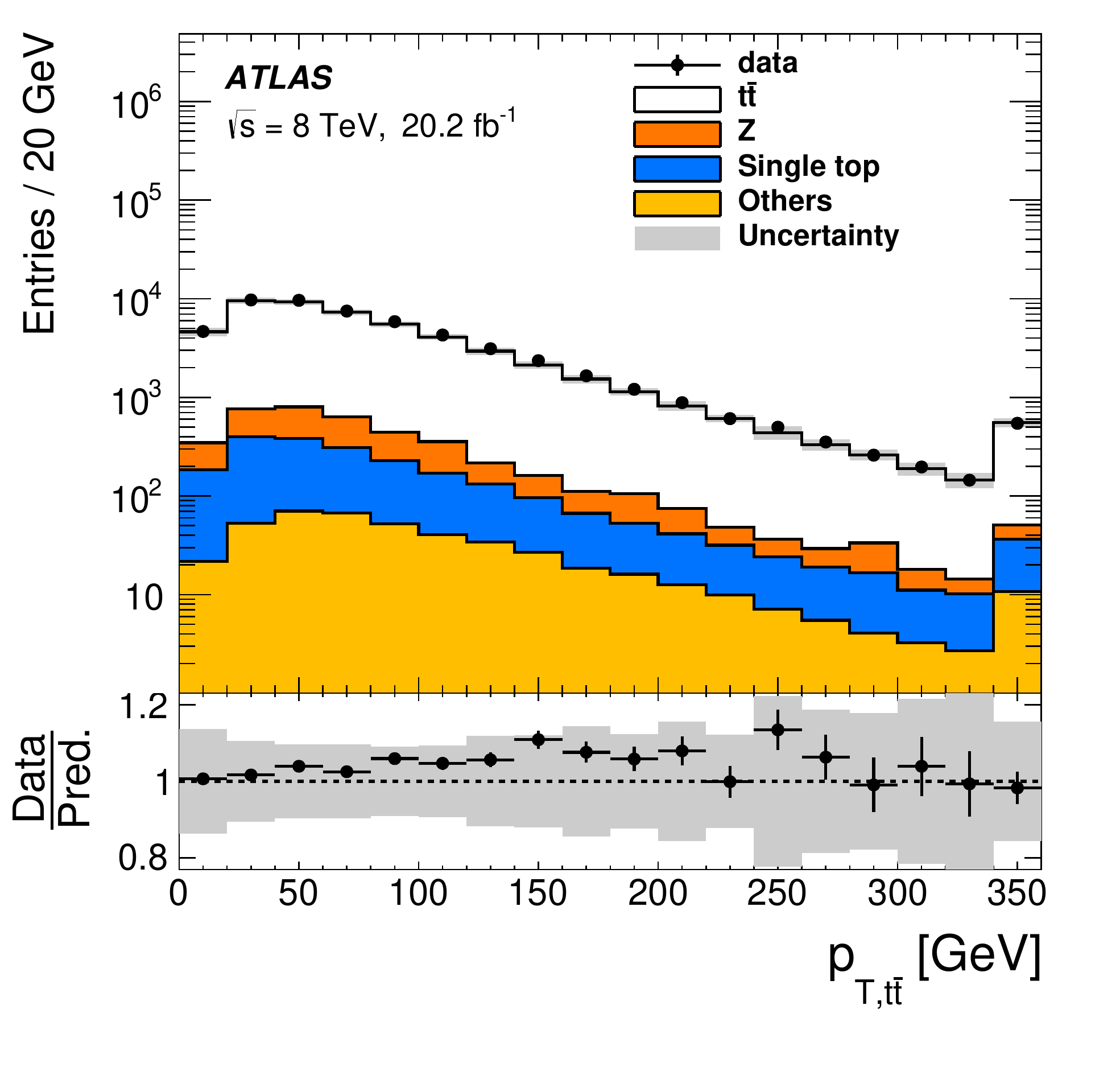}
   \includegraphics[scale=0.33]{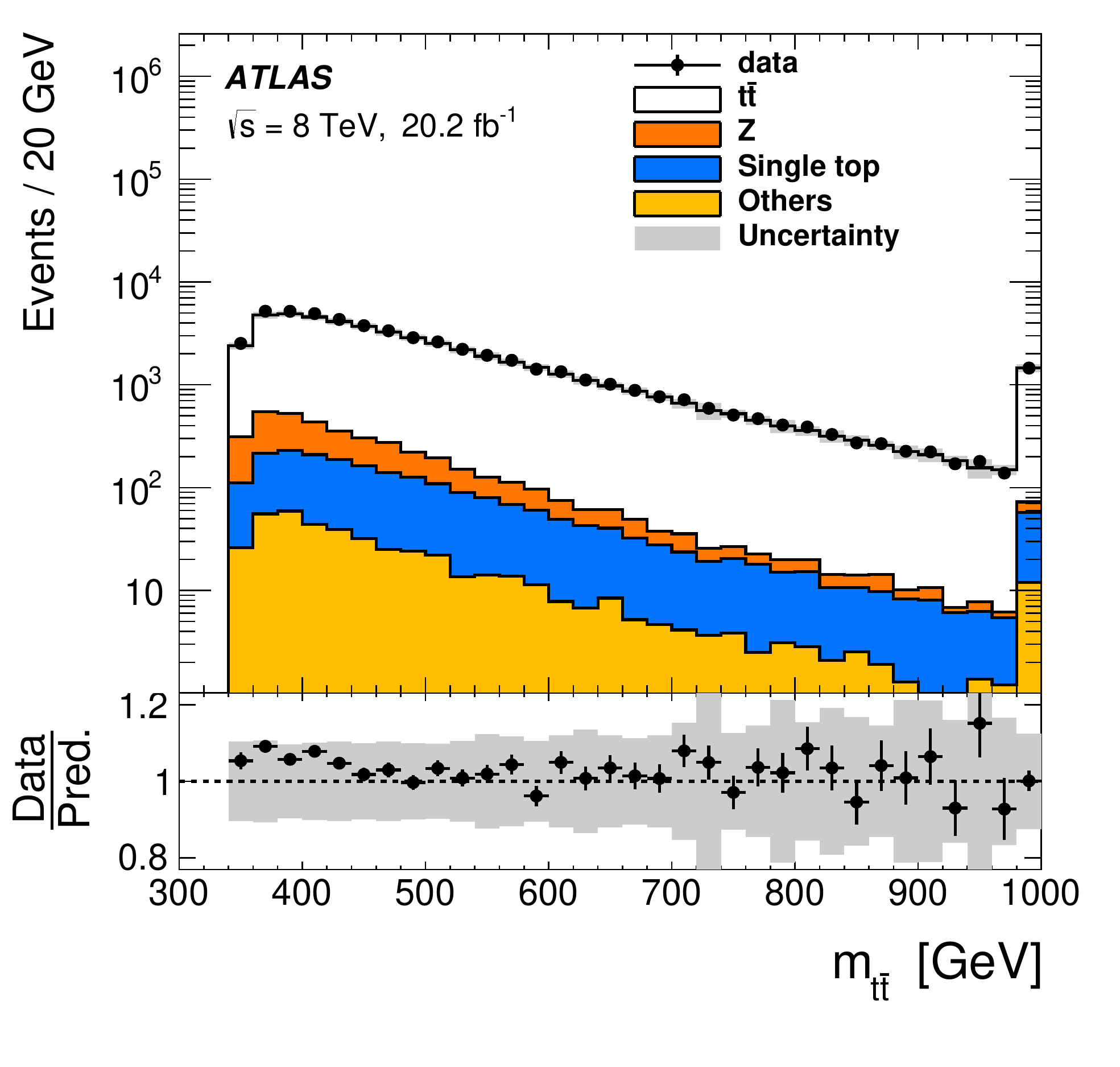}
   \caption{Comparison between data and
     predictions after the kinematic reconstruction in the combined dilepton
     channel. The distributions of the top quark \pt and $\eta$ are shown, as well as the \ttbar \pt and mass.
     The ratio between the data and prediction is also shown. The grey
     area shows the statistical and systematic uncertainty on the signal and background. 
     The $t\bar{t}V$, diboson and fake-lepton backgrounds are shown together in the "Others" category.
     The last bin of the distribution corresponds to the overflow.} 
   \label{fig:reco_dataMC}
 \end{center}
\end{figure}

Figure~\ref{fig:sel_dataMC} shows the jet
multiplicity, lepton \pt and jet \pt for all three channels. 
Figure~\ref{fig:reco_dataMC} shows kinematic distributions of the top quark
and the \ttbar system after the event reconstruction.
The data are well modelled by the MC predictions.
The corrections to the Drell--Yan and fake-lepton backgrounds are applied.
Only the events passing the kinematic reconstruction are 
considered in the distributions.
The total number of predicted events is slightly lower than the number of observed events, but the two are compatible within the systematic uncertainties.
The measurement is insensitive to a difference of normalisation of the signal.
There is also a slight slope in the ratio between data and prediction for the lepton and top quark \pt distributions. 
This is related to a known issue in the modelling of the top quark \pt, described in Section~\ref{sec:modellingsysts}.

The final yields for each channel as well as for the inclusive channel
combining \eeme, \emume and \mumume, along with their combined statistical and systematic uncertainties, can be found in
Table~\ref{tab:event_yields}. The predictions agree with data within 
uncertainties in all channels.

\begin{table}
  \begin{center}
    \renewcommand{\arraystretch}{1.1}
    \sisetup{retain-explicit-plus}
\begin{tabular}{
 l 
 r@{$\,\,$}
 c@{$\,\,$}
 r@{$\,\,\,\,\,\,$}
 r@{$\,\,$}
 c@{$\,\,$}
 r@{$\,\,\,\,\,\,$}
 r@{$\,\,$}
 c@{$\,\,$}
 r@{$\,\,\,\,\,\,$}
 r@{$\,\,$}
 c@{$\,\,$}
 r
}

\hline
               Channel         & \multicolumn{3}{c}{$ee$}&
               \multicolumn{3}{c} {$e\mu$} &
               \multicolumn{3}{c}{$\mu\mu$ } &
               \multicolumn{3}{c}{$\ell \ell$}     \\
\hline

$t\bar{t}$                 & 9140    &$\pm$  & 730  & 27400 & $\pm$ &
1900   & 10800 &$\pm$ & 710 & 47340 & $\pm$ & 2160\\
Fakes                      & 47      &$\pm$  & 78   & 126   & $\pm$ &
62     & 42    &$\pm$ & 32 & 215 & $\pm$ & 105\\
Single-top                 & 342     &$\pm$  & 34   & 1024  & $\pm$ &
85     & 396   &$\pm$ & 35 & 1762 & $\pm$ & 98\\
Diboson                    & 17      & $\pm$ & 4    & 47    & $\pm$ &
7      & 25    &$\pm$ & 4 & 89 & $\pm$ & 9\\
$t\bar{t}V$                & 32      &$\pm$  & 35   & 82    & $\pm$ &
85     & 35    &$\pm$ & 38 & 149 & $\pm$ & 99\\
$Z \rightarrow ee $        & 910     &$\pm$  & 200  &
\multicolumn{3}{c}{--} & \multicolumn{3}{c}{--} & 910 & $\pm$ &200\\
$Z \rightarrow \mu\mu $    & \multicolumn{3}{c}{--} &
\multicolumn{3}{c}{--} & 1100  &$\pm$ & 230 & 1100 & $\pm$  &230\\
$Z \rightarrow \tau\tau $  & 25      &$\pm$  & 10  & 93    & $\pm$ &
20     & 46    &$\pm$ & 15 & 164 & $\pm$ & 27\\
\hline
Total Expected             & 10510   &$\pm$  & 760  & 28800 &$\pm$ &
1900    & 12460 &$\pm$ & 750 & 51750 & $\pm$ & 2180\\
\hline
Data                       & \multicolumn{3}{c}{11162}
&\multicolumn{3}{c}{29985} &\multicolumn{3}{c}{12430} &
\multicolumn{3}{c}{53577} \\

\hline
\end{tabular}

    \caption{Event yields of \ttbar signal, background processes and
      data after the full event selection and the kinematic reconstruction. 
      The given uncertainties
      correspond to the combination of statistical and systematic uncertainties of the
      individual processes.  The last column represents the inclusive
      dilepton channel.}
    \label{tab:event_yields}
  \end{center}
\end{table}

\section{Analysis}
\label{sec:analysis}

Two different measurements of the spin observables are performed.  One
set of measurements is corrected to parton level and the other set is corrected 
to stable-particle level. 
These two levels are defined in the next section, as well as the phase-spaces to which 
the measurements are corrected.

\subsection{Truth level definitions}

\subsubsection{Parton-level definition}
At parton level, the considered top quarks are 
taken from the MC history after radiation but before decay. 
Parton-level leptons include
tau leptons before they decay into an electron or muon and before radiation.
With these definitions, the polarisation can be extracted from the
slope of the \costheta distribution of parton-level particles 
(Equation~(\ref{eqn:single-diff})) and the correlation can be extracted from the mean
value of the distribution (Equation~(\ref{eqn:spin_relation})). 
The measurement corrected to parton level is extrapolated to the full phase-space, where all generated dilepton events are considered.

\subsubsection{Stable-particle definition and fiducial region}
Stable-particle level includes only particles with a lifetime larger than 30~ps.
The charged leptons are required not to originate from hadrons.
Photons within a cone of $\Delta R = 0.1$ around the lepton direction 
are considered as bremsstrahlung and so their four-momenta are added 
to the lepton four-momentum.
Selected leptons are required to have $p_T > 25$~\GeV\ and $|\eta| < 2.5$.
Jets are clustered from all stable particles, excluding the already selected
leptons, by an anti-$k_{t}$ algorithm with a radius parameter $R=0.4$. 
Neutrinos can be clustered within jets.
Intermediate $b$-hadrons have their momentum set to zero, and are allowed to be clustered into the jets 
along with the stable particles~\cite{Cacciari:2008gn}.
If after clustering a $b$-hadron is found in a jet, the jet is considered as $b$-tagged~\cite{Cacciari:2008gn}. 
Jets must have a transverse momentum of at least 25 \GeV\ and have a
pseudorapidity of $|\eta| < 2.5$. Events are rejected if a lepton and a jet are 
separated by $\Delta R < 0.4$.
A fiducial phase-space close to the detector and selection acceptance is defined 
by requiring the presence of at least two leptons and at least two jets
satisfying the kinematic selection criteria. Around 32\% of all generated events
satisfy the fiducial requirements.
No $b$-jet is required in the definition of the fiducial region to keep it common with other analyses not using $b$-tagging. 
The $b$-jets are used in the kinematic reconstruction described in the following.

The top quarks (called pseudo-top-quarks~\cite{Aad:2015eia}) are reconstructed from
the stable particles defined above. If no jets are $b$-tagged, the two highest-\pt
jets are considered for the pseudo-top-quark reconstruction. 
Neutrinos are required not to originate from hadrons, but from $W$ or $Z$
decays or from intermediate tau decays.
For the reconstruction, only
the two neutrinos with the highest \pt are taken in MC events. 
The correct lepton--neutrino pairings are chosen as those with reconstructed masses 
closer to the $W$ boson mass. The correct jet--lepton--neutrino pairings are 
then chosen as those with masses closer to the
generated top quark mass of 172.5 \GeV.

In contrast to the parton-level measurement where all events are included,
   events from outside the fiducial region can still pass the event selection at reconstruction level
   and have to be treated as additional background (called the
       non-fiducial background). This contribution is estimated from 
   background-subtracted data by applying the binwise ratio of non-fiducial to
   total reconstructed signal events obtained from MC simulation, which is found to be
   constant for different levels of polarisation and correlation with
   an average of about 6.5\%.
\subsection{Unfolding}
\label{sec:unfolding}

Selection requirements and detector resolution distort the
reconstructed distributions. An unfolding procedure is applied to correct for these
distortions. The Fully Bayesian Unfolding~\cite{FBU} method is used.
It is based on Bayes' theorem and estimates 
the probability ($p$) of $\Truth{}\in{}\Real{}^{N_t}$ being the true spectrum given the observed
data $\Data\in\Integer^{N_r}$.\footnote{$\Real{}$ and $\Integer$ are the sets of real and natural numbers. $N_{t}$
and $N_{r}$ are the number of bins for the true and reconstructed
distributions.}
This probability is proportional to the likelihood ($\mathcal{L}$) of obtaining the data distribution given
a true spectrum and a response matrix
$\TrasfMatrix\in\Real{}^{N_r}\times{}\Real{}^{N_t}$.
This can be expressed as
\begin{equation}
  \conditionalProb{\Truth{}}{\Data{},\TrasfMatrix{}}
  \propto{}
  \conditionalLhood{\Data{}}{\Truth{},\TrasfMatrix{}}
  \cdot{}
  \pi{}\left(\Truth{}\right),
\end{equation}

where $\pi{}$ is the prior probability density for the true spectrum $\Truth{}$ 
and is taken to be uniform. The
background is estimated as described in Section~\ref{sec:bkg} 
and included in the computation of the likelihood by taking into account
its contribution in data when comparing it with the true spectrum.
The response matrix $\TrasfMatrix{}$, in which each entry $\TrasfMatrix{}_{ij}$ gives the
probability of an event generated in bin $i$ to be reconstructed in
bin $j$, is calculated from the nominal
signal sample. 
By taking a rectangular response matrix connecting the
three different analysis channels to the same true spectrum, the
channels are combined within the unfolding method. The unfolded value
is taken to be the mean of the posterior distribution with its root mean square
taken as the uncertainty.

Different systematic uncertainties are estimated within the unfolding
by adding nuisance parameter terms ($\boldsymbol{\theta}$) to the likelihood for each systematic
uncertainty considered. The so-called marginal likelihood is then
defined as
\begin{equation}
\conditionalLhood{\Data{}}{\Truth{}} = \int \conditionalLhood{\Data{}}{\Truth{},\boldsymbol{\theta}}\cdot \pi(\boldsymbol{\theta})\textrm{d}\boldsymbol{\theta},
\end{equation}

where $\pi(\boldsymbol{\theta})$ is the prior probability density 
for each nuisance parameter $\boldsymbol{\theta}$.
They are defined as Gaussian distributions $G$
with a mean of zero and a width of one. Systematic uncertainties can
be distinguished between normalisation-changing uncertainties ($\theta_n$) and uncertainties
changing both the normalisation and the shape ($\theta_s$) of the reconstructed
distribution of signal $R(\Truth{}; \theta_s)$ and background
$B(\theta_s, \theta_n)$. The marginal likelihood can
then be expressed as:
\begin{equation}
\conditionalLhood{\Data{}}{\Truth{}} = \int \conditionalLhood{\Data{}}{R(\Truth{}; \theta_s),B(\theta_s, \theta_b)}\cdot G(\theta_s) \cdot G(\theta_b) \textrm{d}\theta_s \textrm{d}\theta_b.
\end{equation}

\begin{figure}[!htbp]
 \begin{center}
   \includegraphics[width=0.7\linewidth]{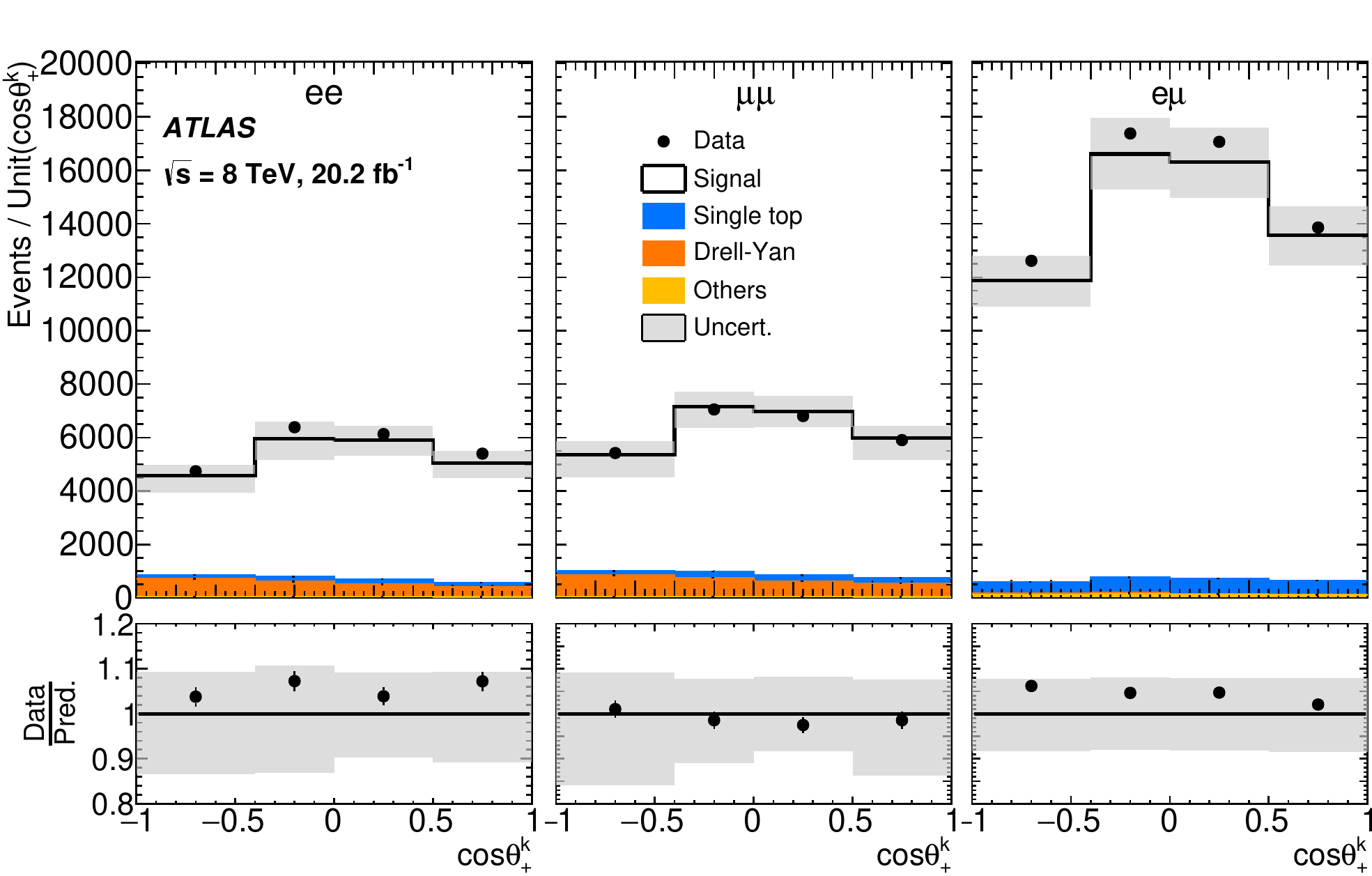}
   \includegraphics[width=0.7\linewidth]{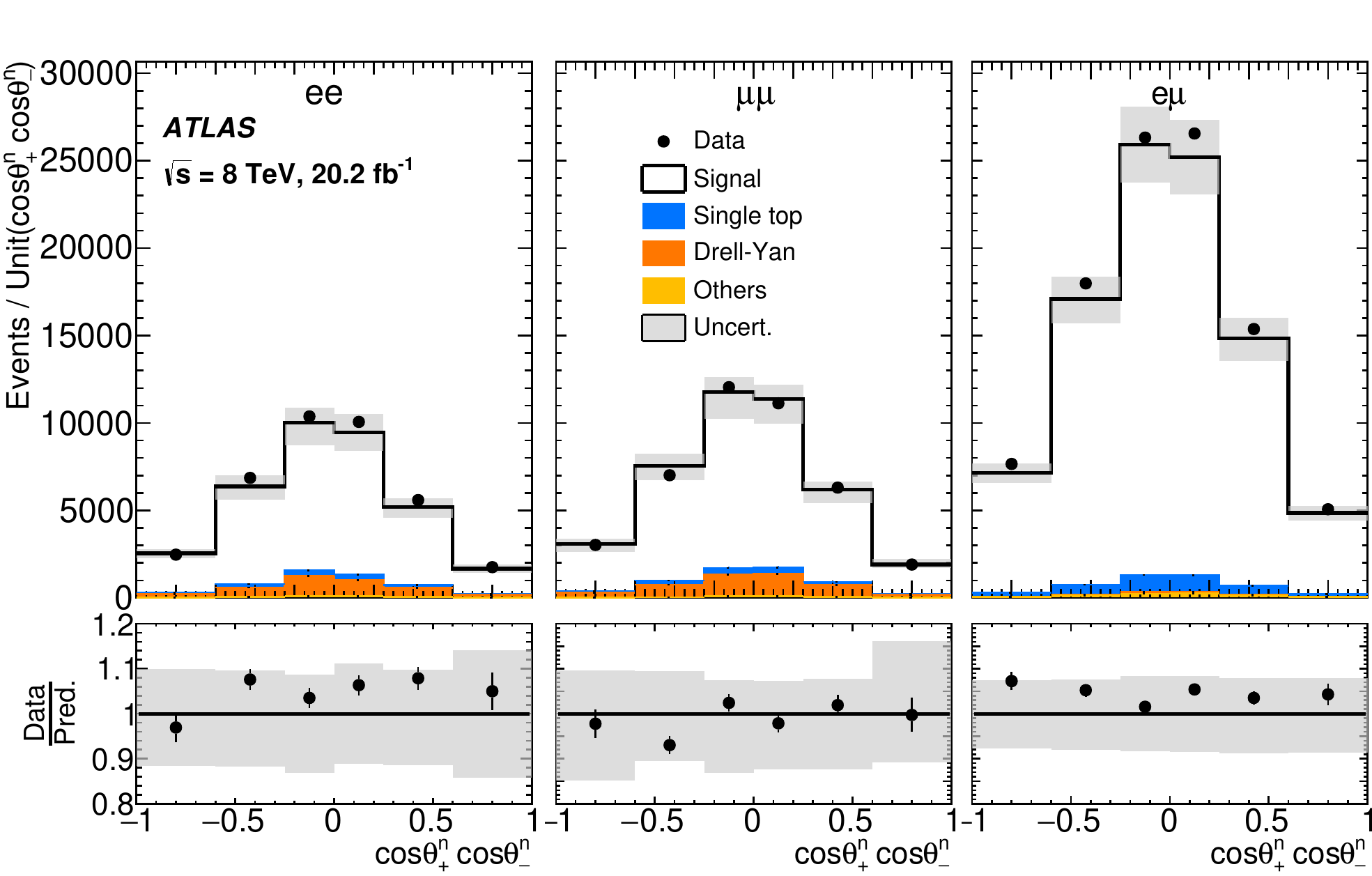}
   \includegraphics[width=0.7\linewidth]{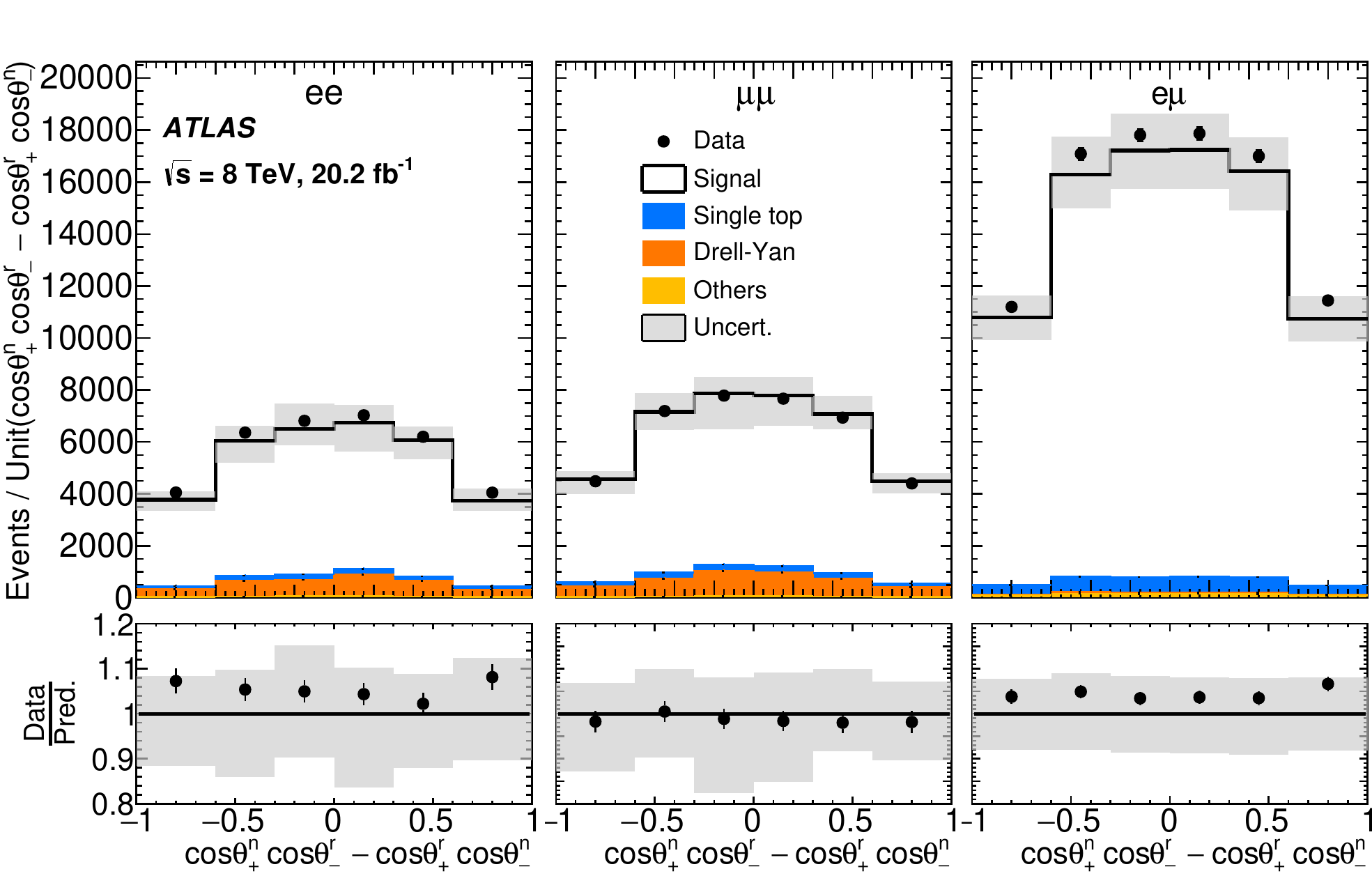}
   \caption{Input distributions for the unfolding procedure of
     \helcosplus, \transcoscos, and \transrcoscosdiff.
     The ratio between the data and prediction is also shown. The grey
     area shows the total uncertainty on the signal and background. 
     The $t\bar{t}V$, diboson and fake-lepton backgrounds are shown together in the "Others" category.} 
   \label{fig:unf_input_dataMC}
 \end{center}
\end{figure}

\begin{figure}[!htbp]
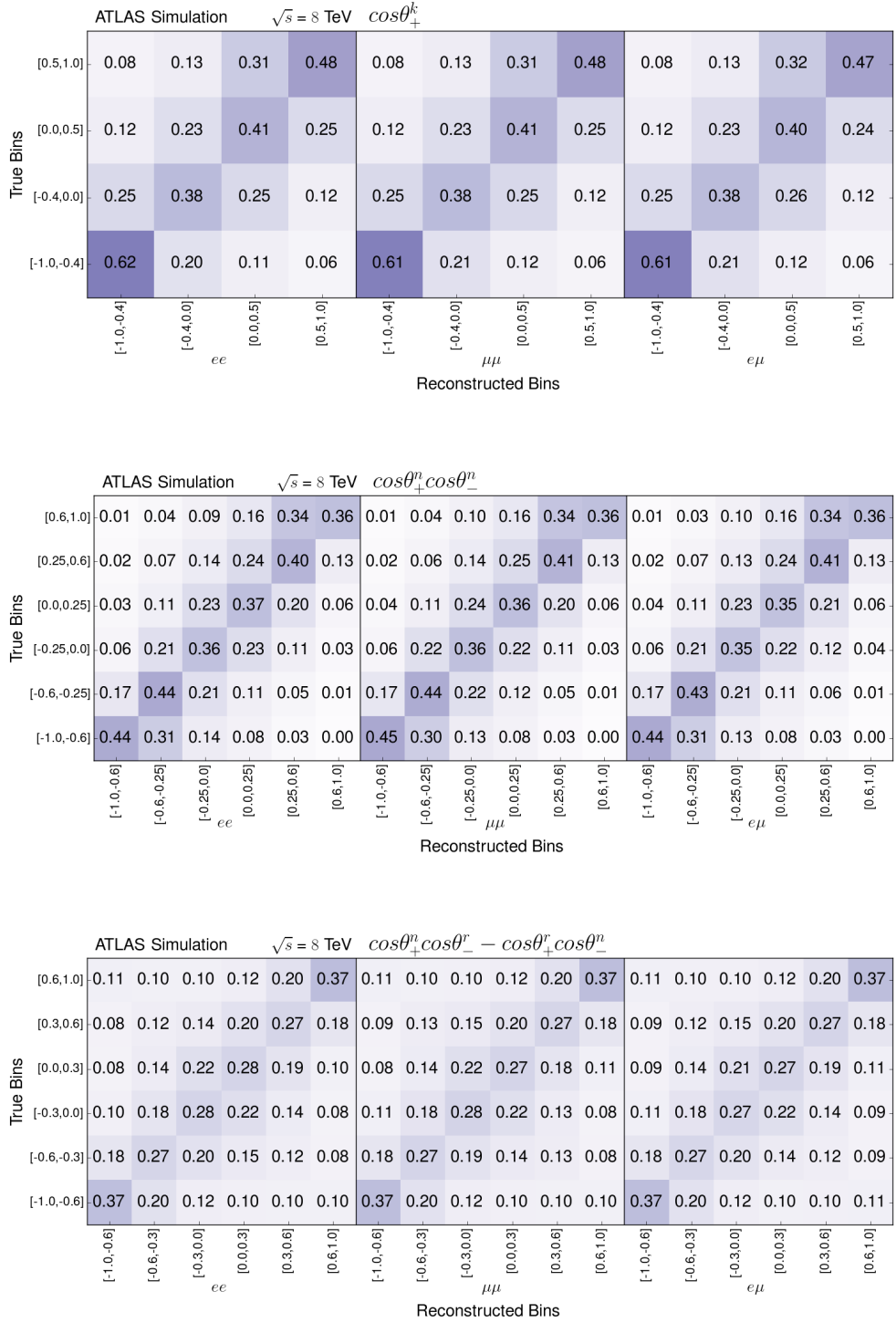

 \begin{center}
   \includegraphics[width=0.8\linewidth]{figures/resmats/resmat_helpol_plus_full.pdf}
   \\[1cm]
   \includegraphics[width=0.8\linewidth]{figures/resmats/resmat_transcorr_full.pdf}
   \\[1cm]
   \includegraphics[width=0.8\linewidth]{figures/resmats/resmat_transrdiff_full.pdf}
   \caption{Response matrices of observables 
     \helcosplus, \transcoscos, and \transrcoscosdiff.
     at parton level. They are divided into $ee$, $\mu\mu$, and $e\mu$ channels. The matrices are normalised per
     truth bin (rows) for each channel separately.} 
   \label{fig:unf_resmat}
 \end{center}
\end{figure}

The method is validated by performing a linearity test in which distributions with 
known values of the polarisation and spin correlations are unfolded.
The distributions of observables are reweighted to inject different values of
the polarisations and correlations. 
For the polarisations and
spin correlations, the double-differential cross section (Equation~(\ref{eqn:double-diff})) is
used, while a linear reweighting is used for the cross
correlations.
The unfolded value for each reweighted distribution 
is then compared to the true value of polarisation or spin correlation and a calibration curve is built.
Non-closure in the linearity test appears as a slope different from one in the
calibration curve.
The number of bins and the bin widths for each observable are chosen based on its resolution
and optimised by evaluating the expected statistical uncertainty and by limiting the bias 
in the linearity test.
The binning optimisation leads to a four-bin configuration
for the polarisation observables and six-bin configurations for the
different correlation observables.
An uncertainty is added to cover the non-closure of the linearity test, which is at most
10\%.
The input distribution and the response matrix normalised per true bin
are shown for one example of polarisation, spin
correlation, and cross correlation in Figures~\ref{fig:unf_input_dataMC} and ~\ref{fig:unf_resmat}. 

\subsection{Systematic uncertainties}
\label{sec:systematics}

The measurement of the spin observables is affected in various ways by
systematic uncertainties. Three different types of systematic uncertainties 
are considered: detector modelling uncertainties affecting both the signal and
background, normalisation uncertainties of the background, and modelling
uncertainties of the signal. The
first two types are included in the marginalisation procedure. 
The reconstructed distribution, varied to reflect a systematic uncertainty,
is compared to the nominal distribution and the average change per bin 
is taken as the width of the Gaussian prior, as discussed in
Section~\ref{sec:unfolding}. 
In order to estimate the impact of each source of systematic uncertainty
individually, pseudodata corresponding to the sum of the nominal signal and
background samples are used. The unfolding procedure with marginalisation
is applied to the pseudodata and constraints on the systematic uncertainties are
obtained. The strongest constraint is on the uncertainty related to the electron identification 
and it reduces this systematic uncertainty by 50\%. The other constraints are of the order of a few percent.
The constrained systematic uncertainties are then used to build the $\pm$1$\sigma$
variations of the prediction. The varied pseudodata are then unfolded
without marginalisation. The impact of each systematic uncertainty is computed by taking half of the
difference between the results obtained from the $\pm$1$\sigma$ variations of pseudodata.

Modelling systematic uncertainties for the signal process are estimated separately by building calibration curves for each sample.
The unfolded value in data is calibrated to generator level using the calibration curves for the nominal sample and the sample varied to reflect the uncertainty.
The difference is taken as the systematic uncertainty.

\subsubsection{Detector modelling uncertainties}
All sources of detector modelling uncertainty are discussed below.

  \subsubsection*{Lepton-related uncertainties}

  \begin{itemize}

  \item{ \textbf{Reconstruction, identification and trigger.}
      The reconstruction and identification efficiencies for electrons and muons, as well as the efficiency of the
      triggers used to record the events differ between data and simulation. Scale factors and their uncertainties are derived using
      tag-and-probe techniques on $Z \rightarrow \ell^{+} \ell^{-} (\ell = e~\text{or}~\mu)$ events in data and in simulated samples to correct the simulation for these differences~\cite{atlas_trigger_perf,atlas_elec_sf}.
    }
  \item{ \textbf{Momentum scale and resolution}. The accuracy of the lepton
    momentum scale and resolution in simulation is checked using
      the  $Z \rightarrow \ell^{+} \ell^{-}$ and $J/\Psi \rightarrow
      \ell^{+} \ell^{-} $ invariant mass distributions. In the case of electrons, $E/p$ studies using
      $W \rightarrow e\nu$ events are also used. Small differences are observed
      between data and simulation. Corrections to the lepton energy scale and
      resolution, and their related uncertainties are also
      considered~\cite{atlas_trigger_perf,atlas_elec_sf}.
    }
  \end{itemize}
  
  \subsubsection*{Jet-related uncertainties}
  
  \begin{itemize}
  \item{ \textbf{Reconstruction efficiency}.
      The jet reconstruction efficiency is found to be about $0.2\%$ lower in the simulation than in data for jets
      below 30~\GeV\ and it is consistent with data for higher jet \pt. To evaluate the systematic uncertainty due to
      this small inefficiency, $0.2\%$ of the jets with \pt below 30~\GeV\ are removed randomly and all jet-related
      kinematic variables (including the missing transverse momentum) are recomputed. The event selection is
repeated using the modified number of jets.
}
\item{ \textbf{Vertex fraction efficiency}.
    The per-jet efficiency to satisfy the jet vertex fraction requirement is measured in $Z \rightarrow \ell^{+} \ell^{-}+\textrm{1-jet}$ events
    in data and simulation, selecting separately events enriched in hard-scatter jets and events enriched in
    jets from pile-up. The corresponding uncertainty
    is estimated by changing the nominal JVF requirement value and
    repeating the analysis using the modified value.
  }
\item{ \textbf{Energy scale}. The jet energy scale (JES) and its uncertainty were derived by combining information from test-beam
    data, LHC collision data and simulation~\cite{atlas_jes}. The jet energy scale uncertainty is split
    into 22 uncorrelated sources, which have different jet $p_\text{T}$ and $\eta$ dependencies
    and are treated independently.
  }
  
\item{\textbf{Energy resolution}. The jet energy resolution was measured
  separately for data and simulation.  A systematic
    uncertainty is defined as the difference in quadrature between the jet energy
    resolutions for data and simulation. To estimate the corresponding
    systematic uncertainty, the jet energy in 
    simulation is smeared by this residual difference. 
}

\item{\textbf{$b$-tagging/mistag efficiency}. Efficiencies to tag jets from $b$- and $c$-quarks in the simulation are
    corrected by \pt- and $\eta$-dependent data/MC scale factors.
    The uncertainties on these scale factors are about 2\% for the
    efficiency for $b$-jets, between 8\% and 15\% for $c$-jets, and between 15\%
    and 43\% for light jets~\cite{atlas_btagging1,atlas_btagging2}. 
  }
\end{itemize}

The dominant uncertainties in this category are related to lepton reconstruction, identification and trigger, jet energy scale and jet energy resolution.
The contribution from this category to the total uncertainty is small (less than 20\% for all observables).

\subsubsection*{Missing transverse momentum}

The systematic uncertainties associated with the momenta and energies
of reconstructed objects (leptons and jets) are propagated to the $\met$ calculation. The $\met$ reconstruction
also receives contributions from the presence of low-\pt jets and
calorimeter energy deposits not included in reconstructed objects (the ``soft term''). 
The systematic uncertainty associated with the soft term is estimated using
$Z \rightarrow \mu^{+} \mu^{-}$ events using methods similar to those used in Ref.~\cite{atlas_met2}.
The effect of this procedure on the measured observables is minor.

\subsubsection{Background-related uncertainties}

The uncertainties on the single-top-quark, $t\bar{t}V$, and diboson backgrounds are 6.8\%, 10\%, and 5\%, respectively~\cite{Kidonakis:2012db,Campbell:2012dh,Garzelli:2012bn}. 
These correspond to the uncertainties on the theoretical cross
sections used to normalise the MC simulated samples.

The uncertainty on the normalisation of the fake-lepton background is
estimated by using various MC generators for each process
contributing to this background. The scale factor in the control region is recomputed for each variation and the change is propagated to the expected number of events in the signal region.
In the $\mu\mu$ channel, the uncertainty is obtained by comparing a
purely data-driven method based on the measurement of the
efficiencies of real and fake loose leptons,
 and the estimation used in this analysis.   
The resulting relative total uncertainties are 170\% in the $ee$ channel,
77\% in the $\mu\mu$ channel and 49\% in the $e\mu$ channel. 

For the Drell--Yan background the detector modelling uncertainties described previously are propagated to the scale factors 
derived in the control region by recalculating them for all the uncertainties.
An additional uncertainty of 5\% is obtained by varying the control region.

Uncertainties on the shape of the different backgrounds were also estimated but found to be negligible.
This category represents a minor source of uncertainty on the measurements.

\subsubsection{Modelling uncertainties}
\label{sec:modellingsysts}

These systematic uncertainties are estimated using the samples listed in Table~\ref{tab:mc_gens_signal}.

\begin{itemize}
\item {\textbf{Choice of MC generator.} The uncertainty is obtained by comparing samples generated with either the \powheg or the \mcatnlo generator, both interfaced with \herwig. It is one of the dominant uncertainties of the measurement.}
  
\item {\textbf{Parton shower and hadronisation.} This effect is estimated by comparing samples generated with \powheg interfaced either with \pythia or \herwig, and is one of the dominant systematic uncertainties.}
  
\item {\textbf{Initial- and final-state radiations.} The uncertainty associated with the ISR/FSR
    modelling is estimated using \powheg interfaced with \pythia  
    where the parameters of the generation were varied to be compatible with the results of a measurement of $t\bar{t}$ production 
    with a veto on additional jet activity in a central rapidity interval~\cite{atlas_isrfsr}.
    The difference obtained between the two samples is divided by two. 
    This uncertainty is large and even dominant for some of the observables.
}

\item {\textbf{Colour reconnection and underlying event.}}  The
  uncertainties associated with colour reconnection and the underlying event
  are obtained by comparing dedicated samples with a varied 
  colour-reconnection strength and underlying-event activity to a reference
  sample. All samples are generated by \powheg and interfaced with
  \pythia. The reference sample uses the Perugia2012 tune, 
  the colour-reconnection sample uses the Perugia2012loCR tune, and the 
  underlying-event sample uses the Perugia2012mpiHi tune.  This uncertainty is large and even dominant for some of the observables.
  
 \item {\textbf{Parton distribution functions.} PDF uncertainties are obtained by using the error sets of CT10~\cite{ct10}, MWST2008~\cite{mstw}, and NNPDF23~\cite{nnpdf}, 
        and following the recommendations of the PDF4LHC working group~\cite{PDF4LHC}.
        The impact of this uncertainty is small.}
     
\item {\textbf{Top quark \pt modelling.} The top quark \pt spectrum is not satisfactorily modelled in MC simulation~\cite{Aad:2015mbv,Khachatryan:2015oqa}. 
       The impact of the mismodelling is estimated by reweighting the simulation to data and unfolding the different distributions using the nominal response matrix.
       The differences with respect to the nominal values are negligible compared to the other modelling uncertainties. 
       The impact of this mismodelling is thus considered negligible, and no uncertainty is added to the total uncertainty.}

    \item{\textbf{Polarisation and spin correlation.}} The response
      matrices used in the unfolding are calculated using the SM
      polarisation and spin correlation. An uncertainty related to a
      different polarisation and spin correlation is obtained by
      changing their values in the linearity test. In the reweighting
      procedure of the spin correlation observables, the polarisation
      is changed by $\pm 0.5$\%, while for the polarisation
      observables, the spin correlation is changed by $\pm 0.1$. This
      uncertainty cannot be applied to the cross correlation
      observables as no analytic description of these observables is
      available. Instead, a linear reweighting is used, not depending on the
      polarisation or spin correlation along any axis as described in Section~\ref{sec:unfolding}.
    \end{itemize}

The impact of this category is large and can represent up to 85\% of the total uncertainty.
    
    \subsubsection{Other uncertainties}
  \begin{itemize}
  \item{\textbf{Non-closure uncertainties.}} 
  When the calibration curve for the nominal signal \powheg sample is estimated
  a residual slope and a non-zero offset are observed. This bias, introduced by
  the unfolding procedure, is propagated to the measured values. This uncertainty is small compared to the total uncertainty.
  \item{\textbf{MC sample size.}}
  The statistical uncertainty of the nominal signal \powheg sample is estimated by
  performing pseudoexperiments on MC events. The migration matrix is varied within the MC statistical
  uncertainty and the unfolding procedure is repeated. The standard deviation of the unfolded polarisation or spin correlation values 
  is taken as the uncertainty. This uncertainty is small compared to the total uncertainty.
  \item{\textbf{Top quark mass uncertainty.}}
  The top quark mass is assumed to be 172.5 \GeV\ in MC simulation and in the reconstruction method. 
  A variation of this value could have an impact on the measurement.
  To estimate this impact, MC samples with different values of the top quark mass are unfolded with the default response matrix. 
  For each observable, the dependence of the unfolded value on the mass is fitted with a linear function and presented in Section~\ref{sec:results}.
  The slope is then multiplied by the 0.70 \GeV\ uncertainty on the most precise ATLAS top quark mass measurements~\cite{Aaboud:2016igd}.
  The obtained uncertainty is presented in the next section, but it is shown separately and is not included in the total uncertainty.
  \end{itemize}

\section{Results}
\label{sec:results}

\begin{figure}[!htbp]
 \begin{center}
   \includegraphics[scale=0.425]{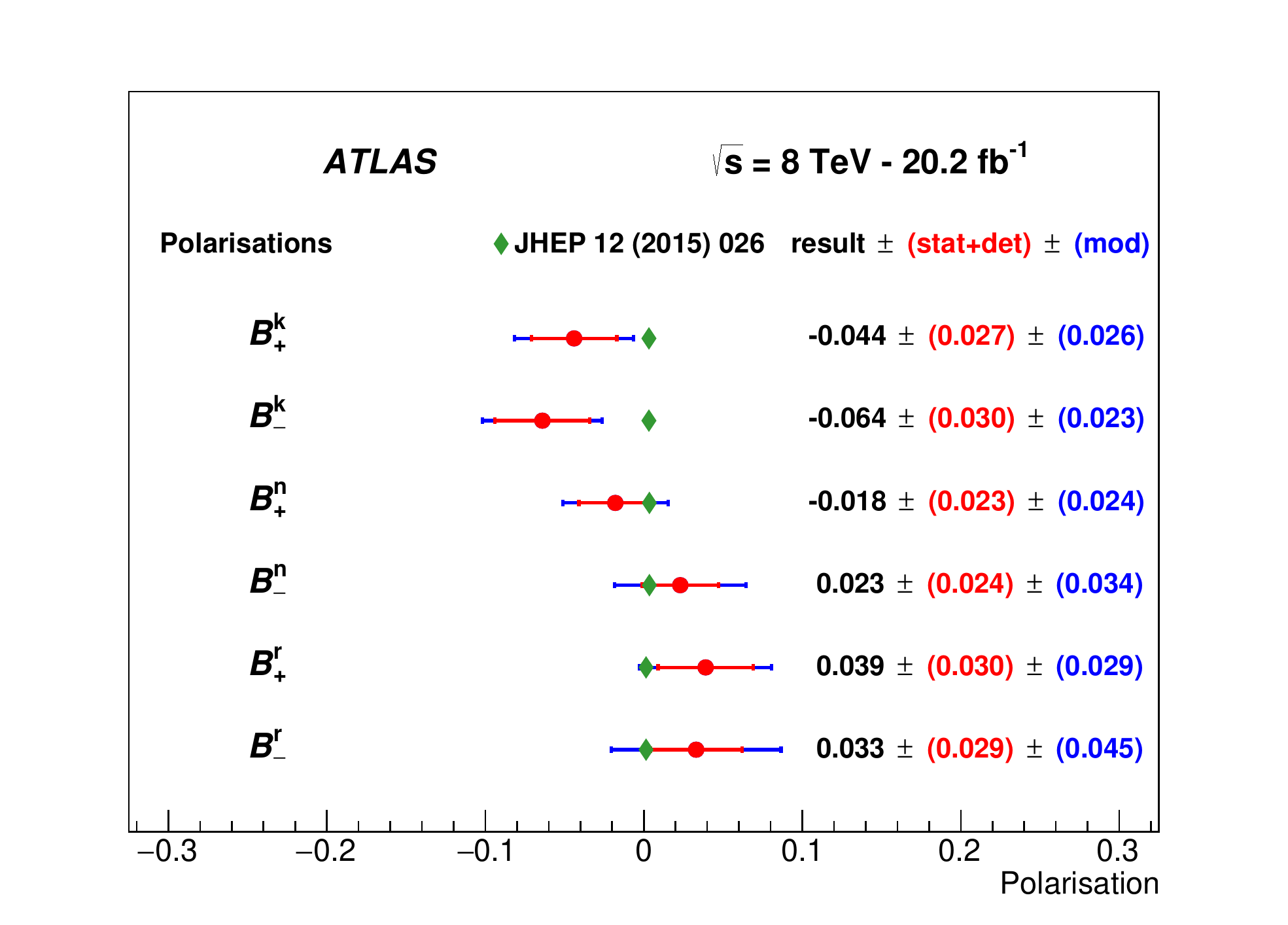}
   \includegraphics[scale=0.425]{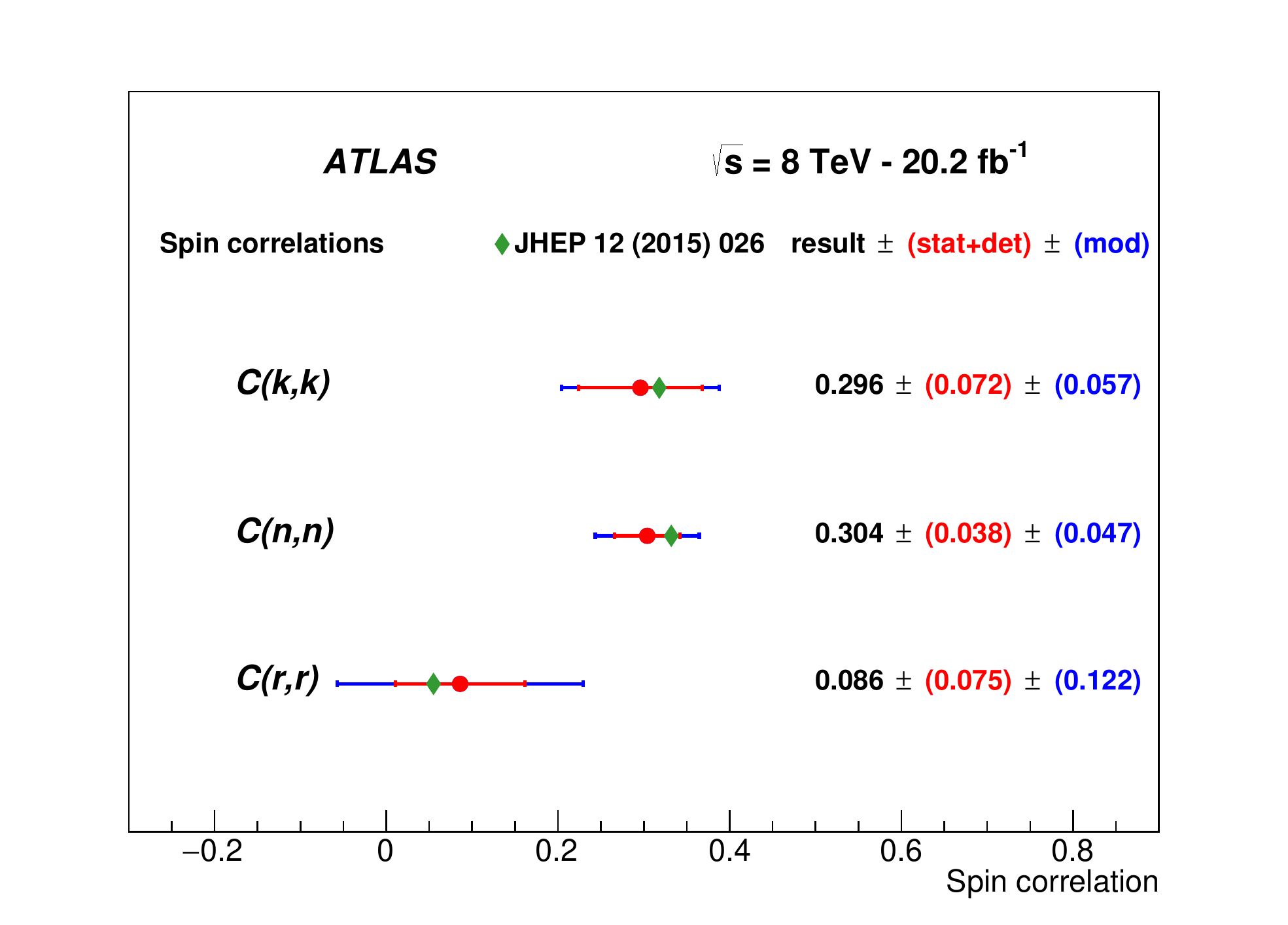}
   \includegraphics[scale=0.425]{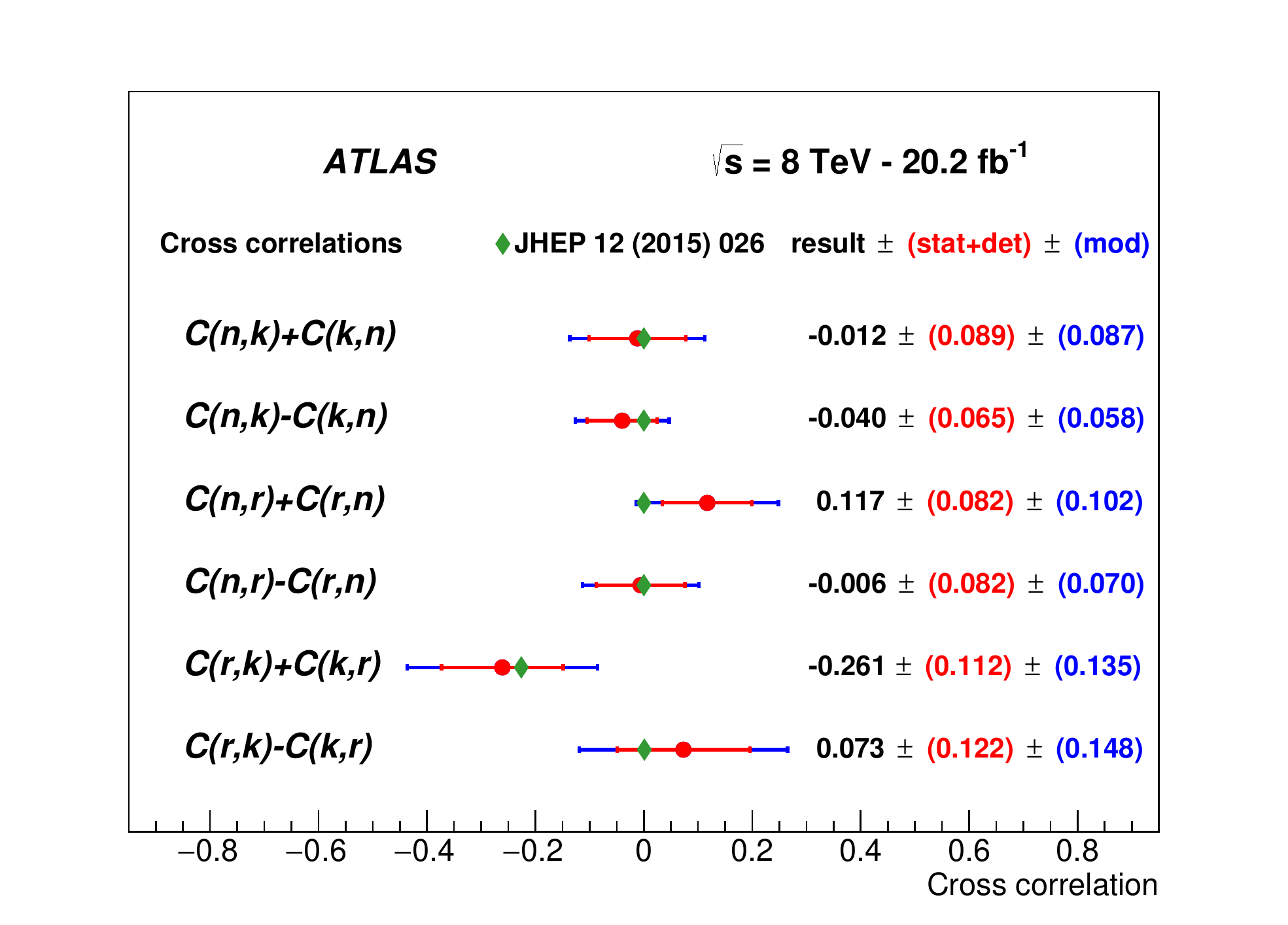}
   \caption{Comparison of the measured polarisations and spin correlations (data points) with predictions
     from the SM (diamonds) for the parton-level measurement. Inner bars indicate uncertainties obtained from the
     marginalisation, outer bars indicate modelling systematics,
     summed in quadrature. 
     The widths of the diamonds are chosen for illustrative purposes only.} 
   \label{fig:results_summary_partonic}
 \end{center}
\end{figure}

Applying the unfolding procedure with marginalisation to the
reconstructed distributions gives the following results at parton and
stable-particle level. Table~\ref{tab:results_summary} presents the results for the
polarisations and correlations at parton level. It shows the central value and the total uncertainty
as well as a breakdown of the systematic uncertainties for the various categories described in 
Section~\ref{sec:systematics}.
Figure~\ref{fig:results_summary_partonic} shows 
the predictions at 8 \TeV\ calculated
in Ref.~\cite{bernreuther_observables} and the unfolded result.
None of the observables deviate significantly from the SM predictions. 
The transverse correlation, \transcorr, 
differs from the case of no spin correlation by
$5.1$ standard deviations. The correlations between the different polarisation and spin correlations were
evaluated and found to be small. The highest correlations are found to be around 10\% 
between the polarisation and spin correlation along the
helicity axis and the $r$-axis and between some cross correlations. 

\begin{figure}[!htbp]
 \begin{center}
   \includegraphics[scale=0.325]{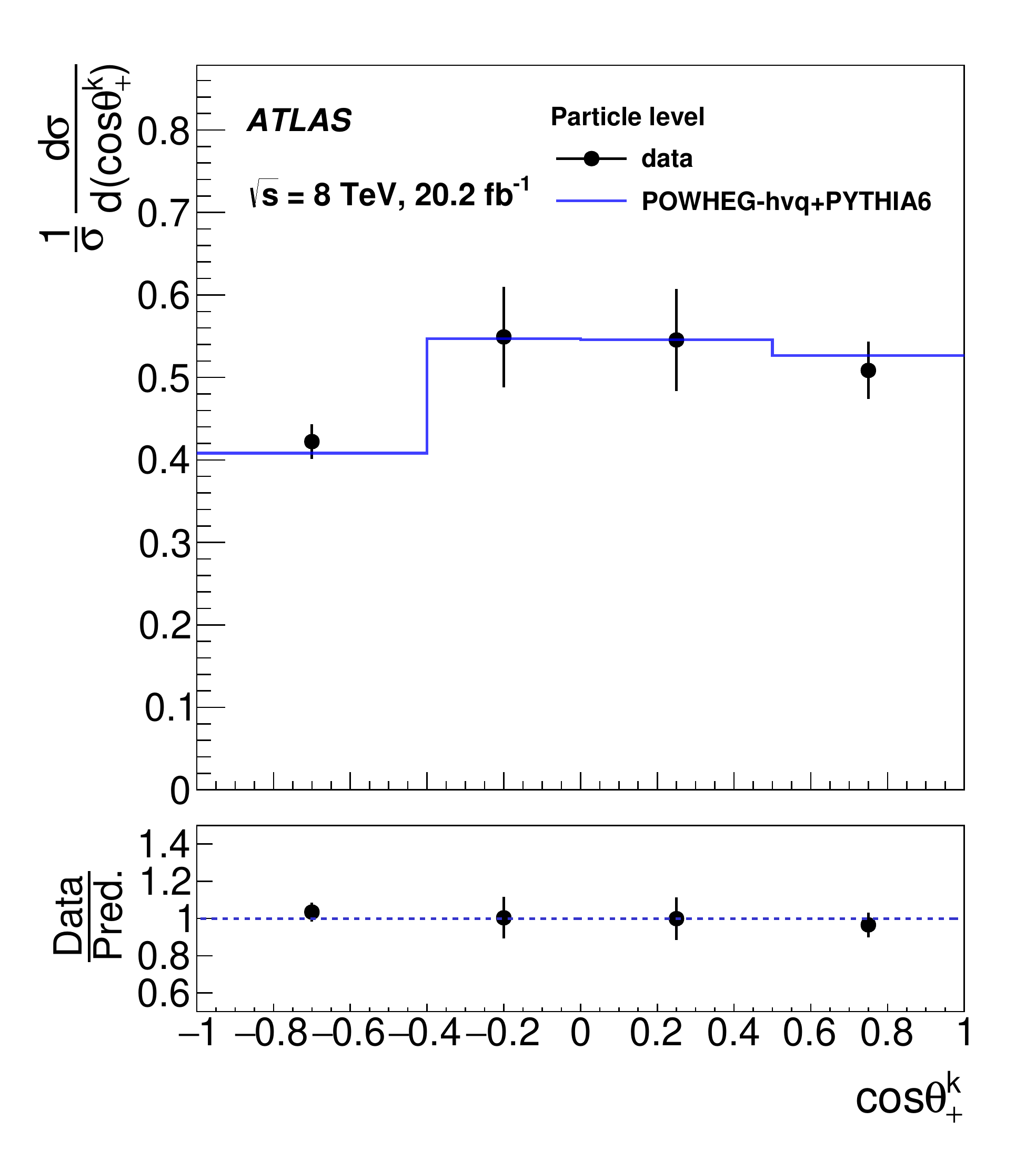}
   \includegraphics[scale=0.325]{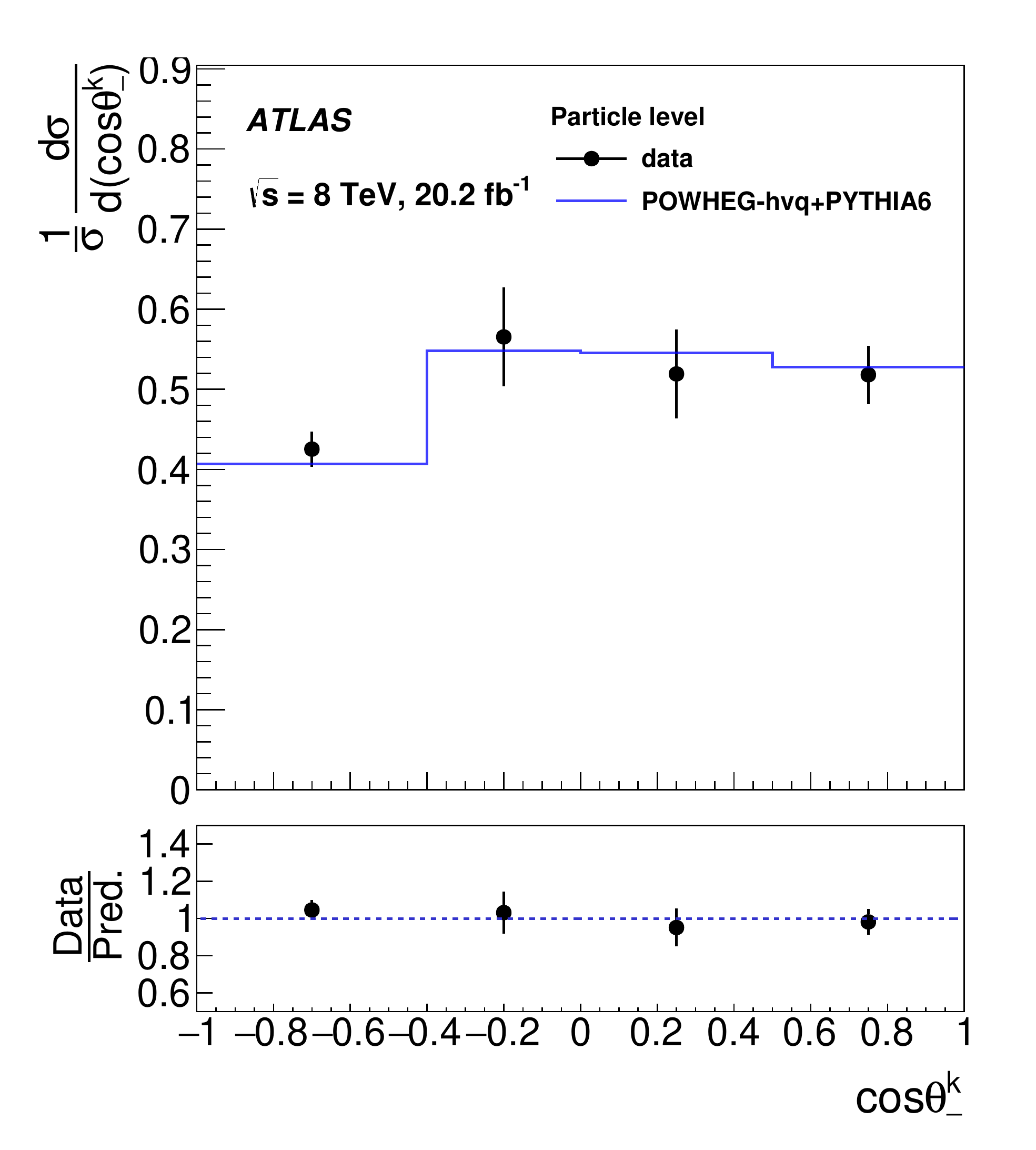} \\
   \includegraphics[scale=0.325]{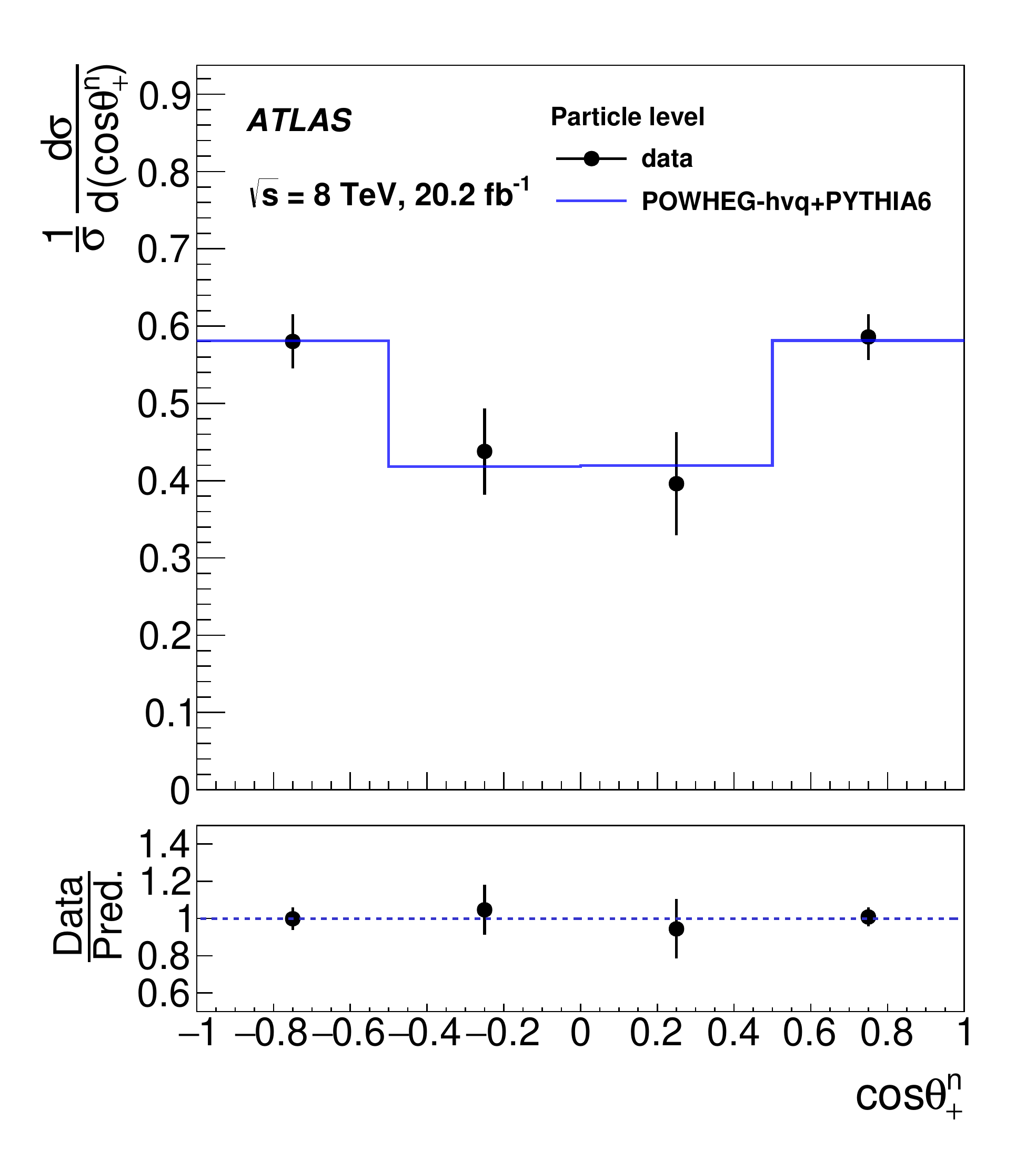}
   \includegraphics[scale=0.325]{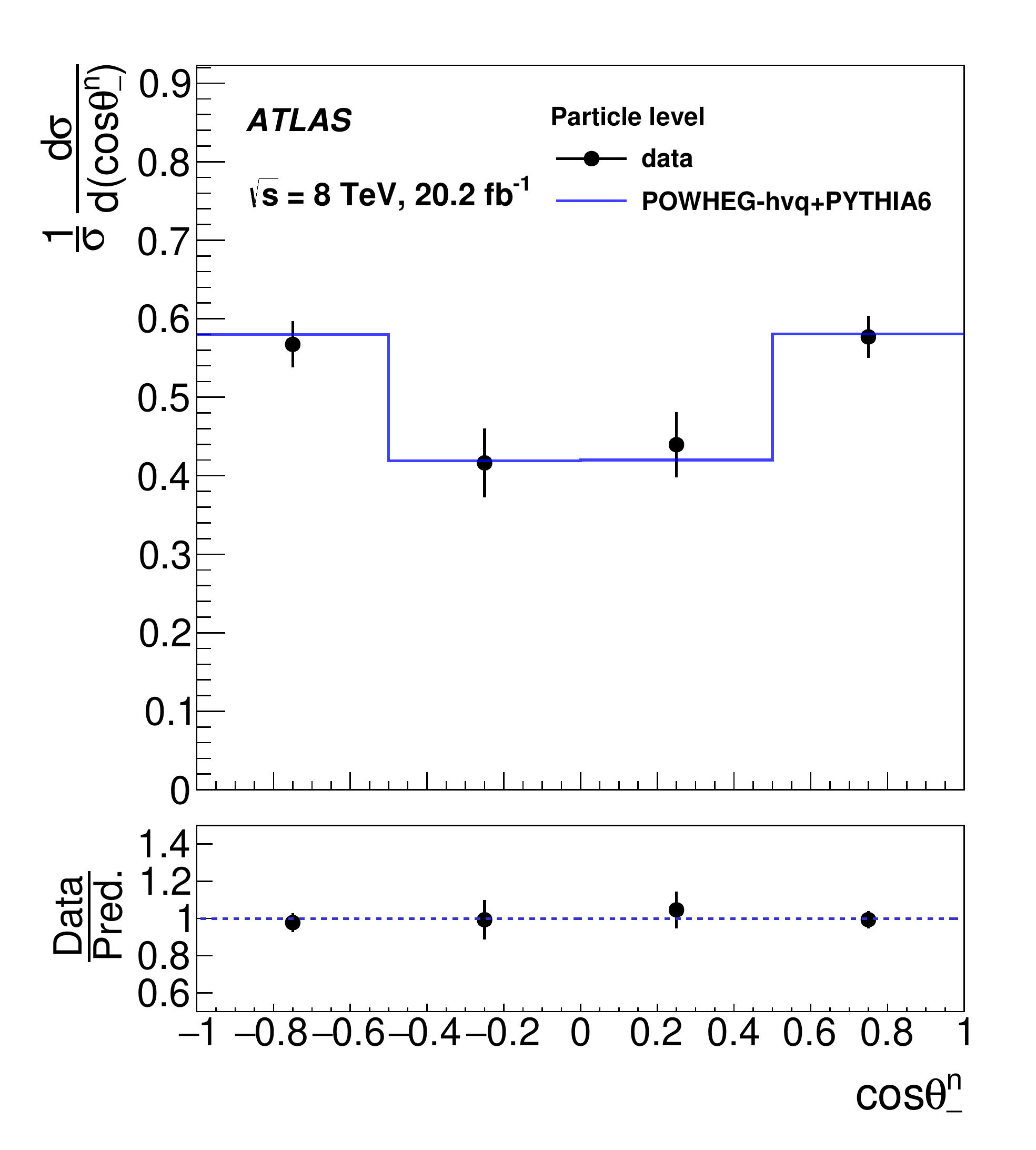} \\
   \includegraphics[scale=0.325]{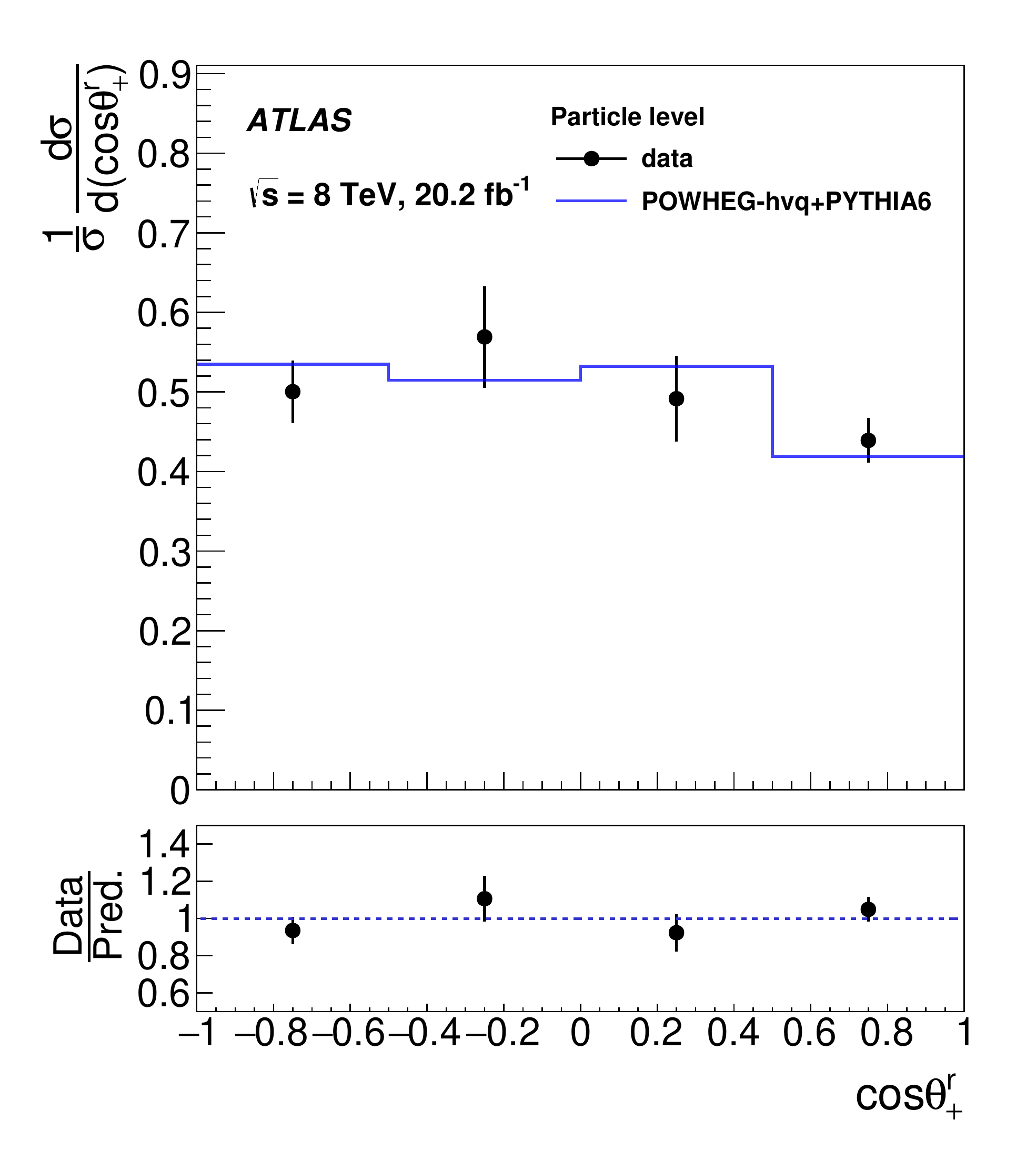}
   \includegraphics[scale=0.325]{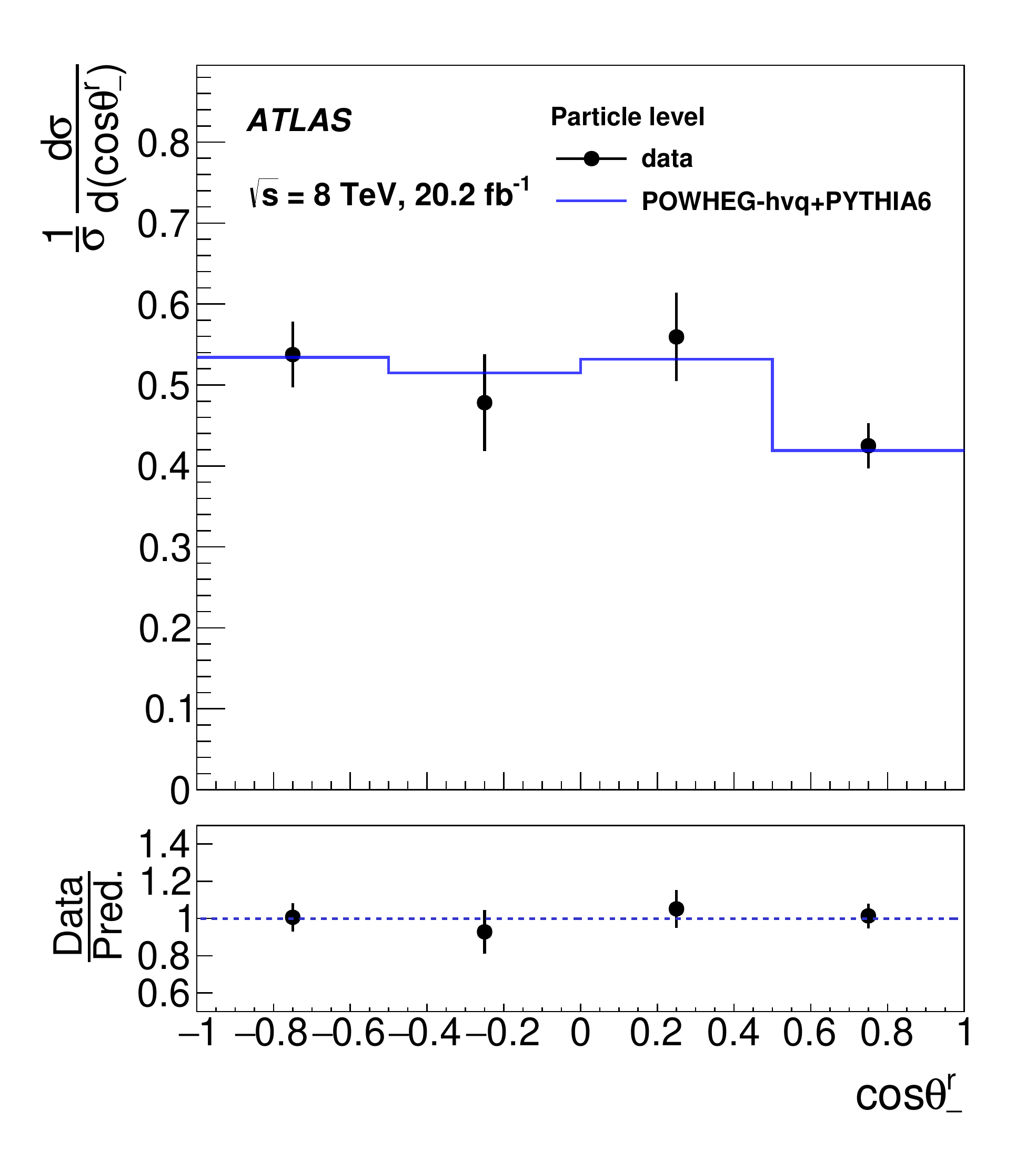} 
   \caption{Comparison of the unfolded polarisation distributions and
     the prediction from the signal MC simulation for the stable-particle
     measurement. The total uncertainty is shown in each bin. The bin-to-bin correlations between adjacent bins are typically between $-0.9$ and $-0.6$. 
     The correlations between non-adjacent bins range from $-0.4$ to 0.6.} 
   \label{fig:results_particle_pol}
 \end{center}
\end{figure}

\begin{figure}[!htbp]
 \begin{center}
   \includegraphics[scale=0.35]{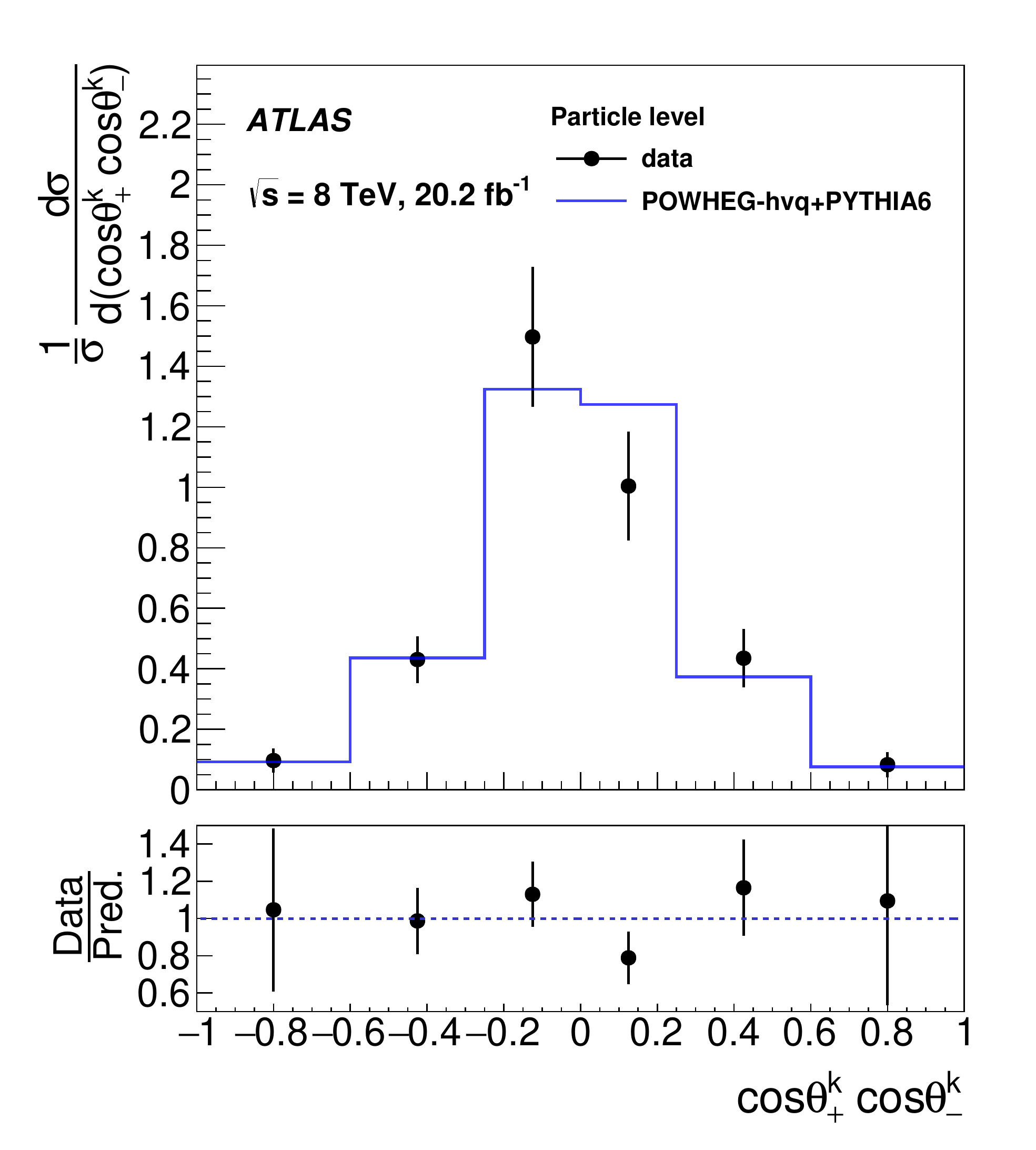}
   \includegraphics[scale=0.35]{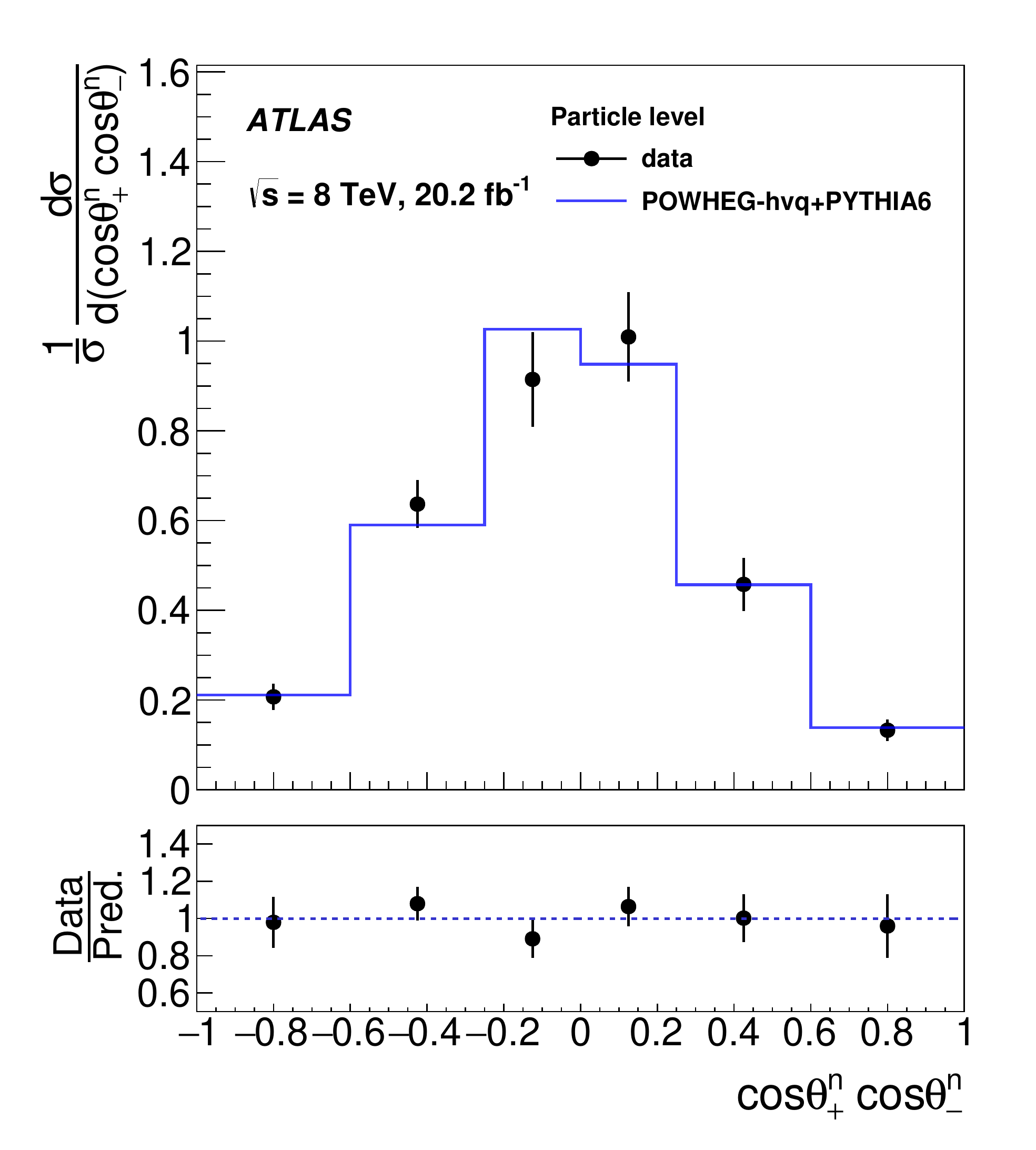} \\
   \includegraphics[scale=0.35]{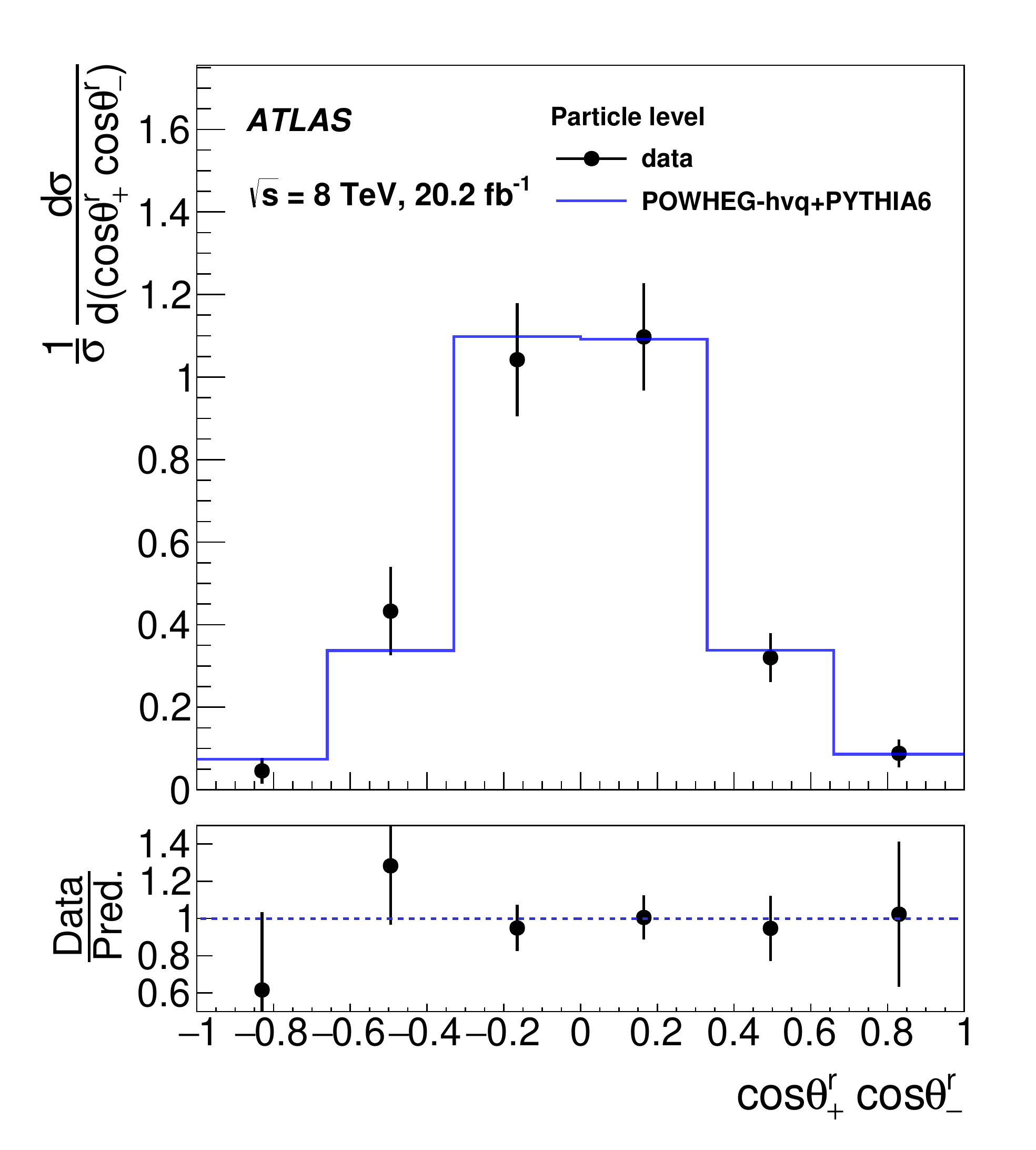}
   \caption{Comparison of the unfolded spin correlation distributions and
     the prediction from the signal MC simulation for the stable-particle
     measurement. The total uncertainty is shown in each bin. 
     The bin-to-bin correlations between adjacent bins are typically between $-0.9$ and $-0.4$. 
     The correlations between non-adjacent bins range from $-0.4$ to 0.6.} 
   \label{fig:results_particle_corr}
 \end{center}
\end{figure}

\begin{figure}[!htbp]
 \begin{center}
   \includegraphics[scale=0.325]{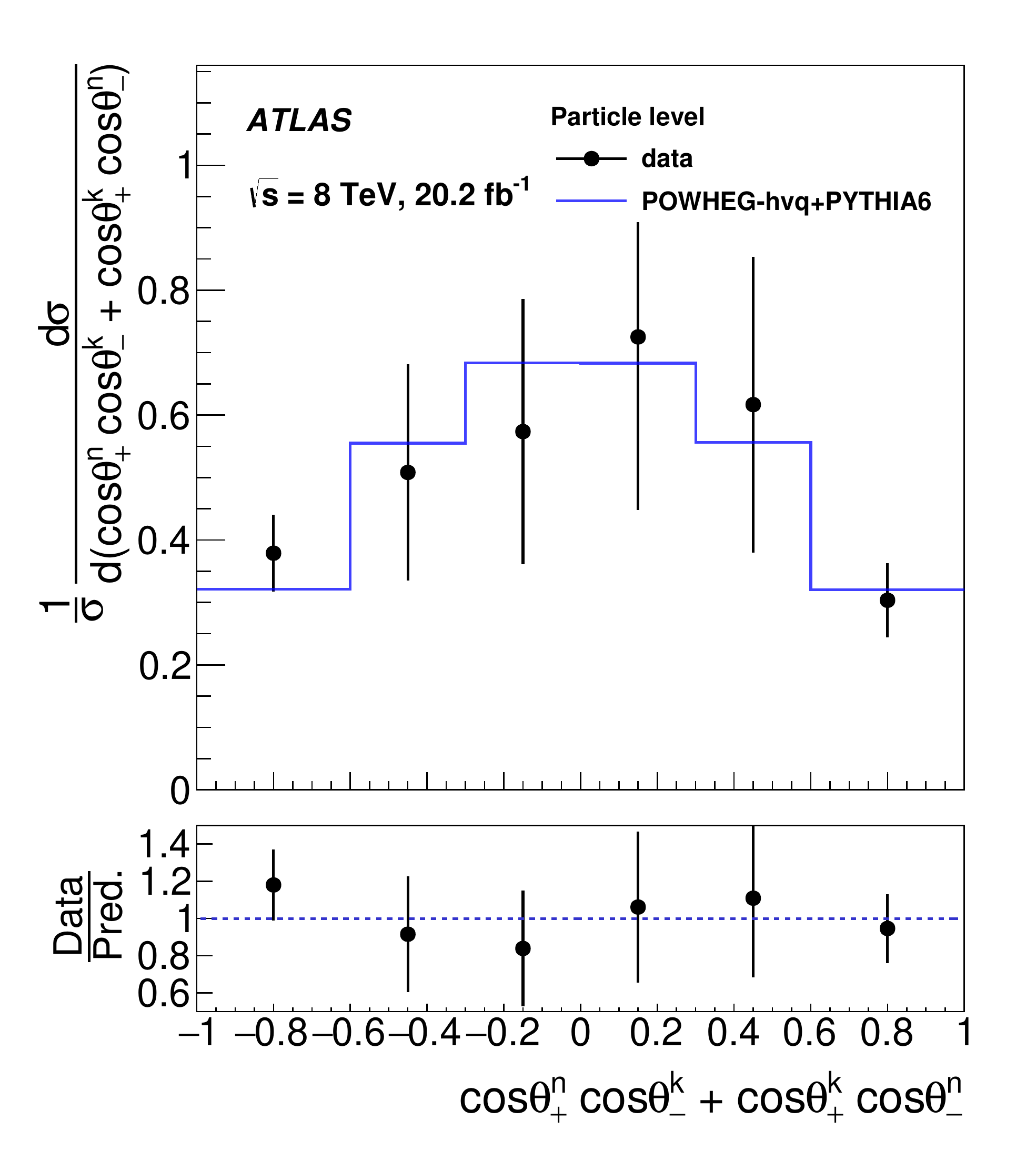}
   \includegraphics[scale=0.325]{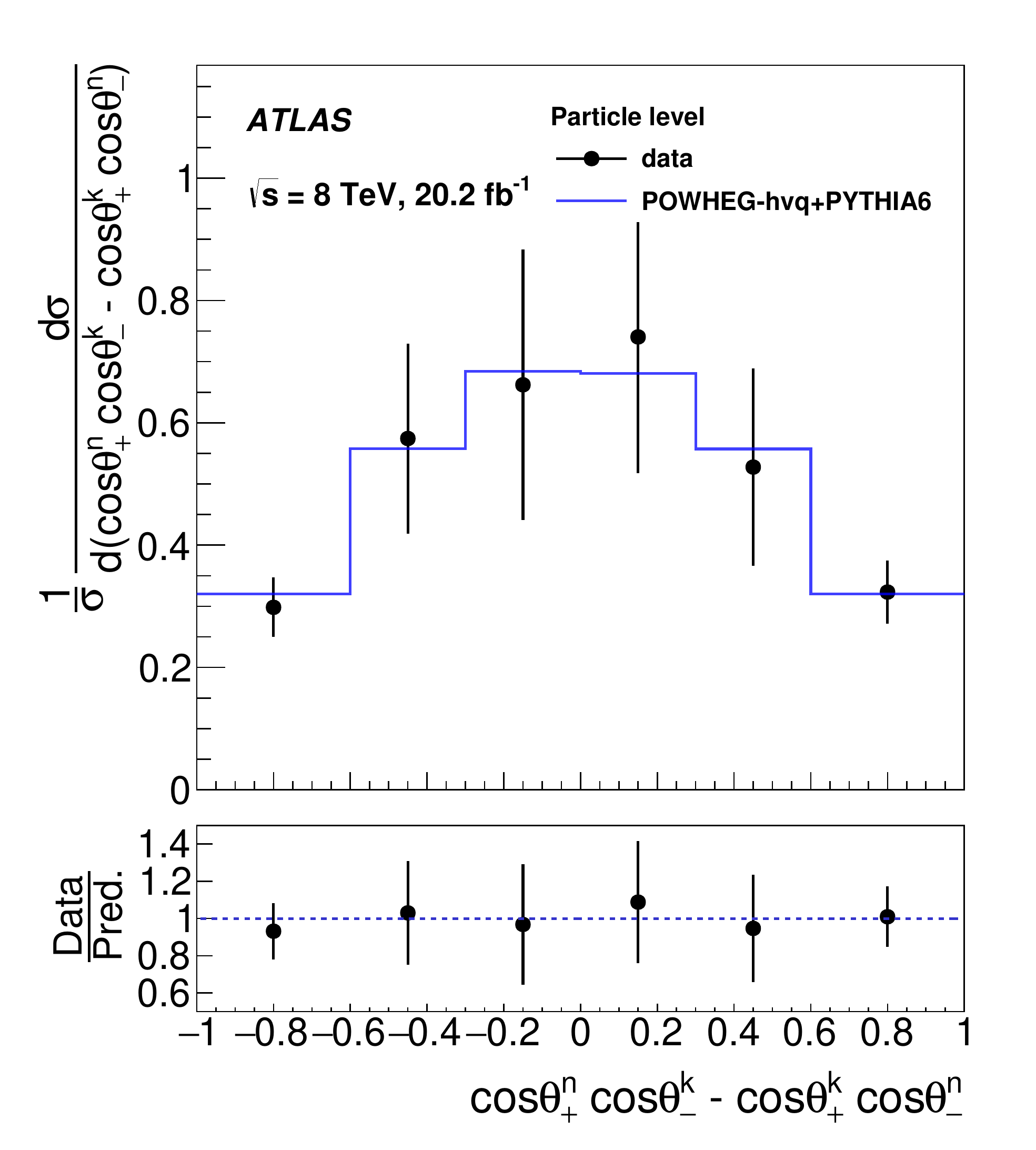} \\
   \includegraphics[scale=0.325]{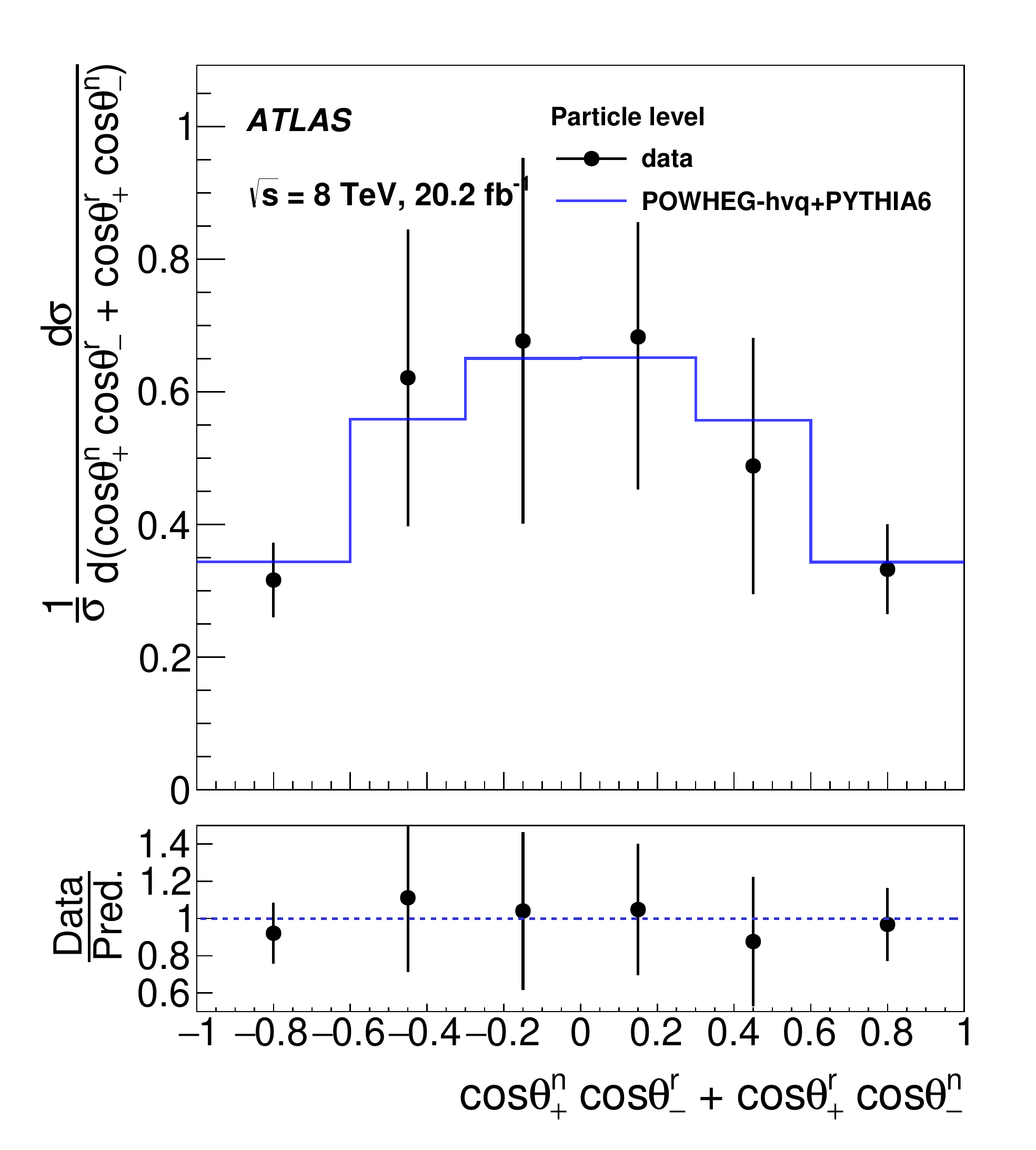}
   \includegraphics[scale=0.325]{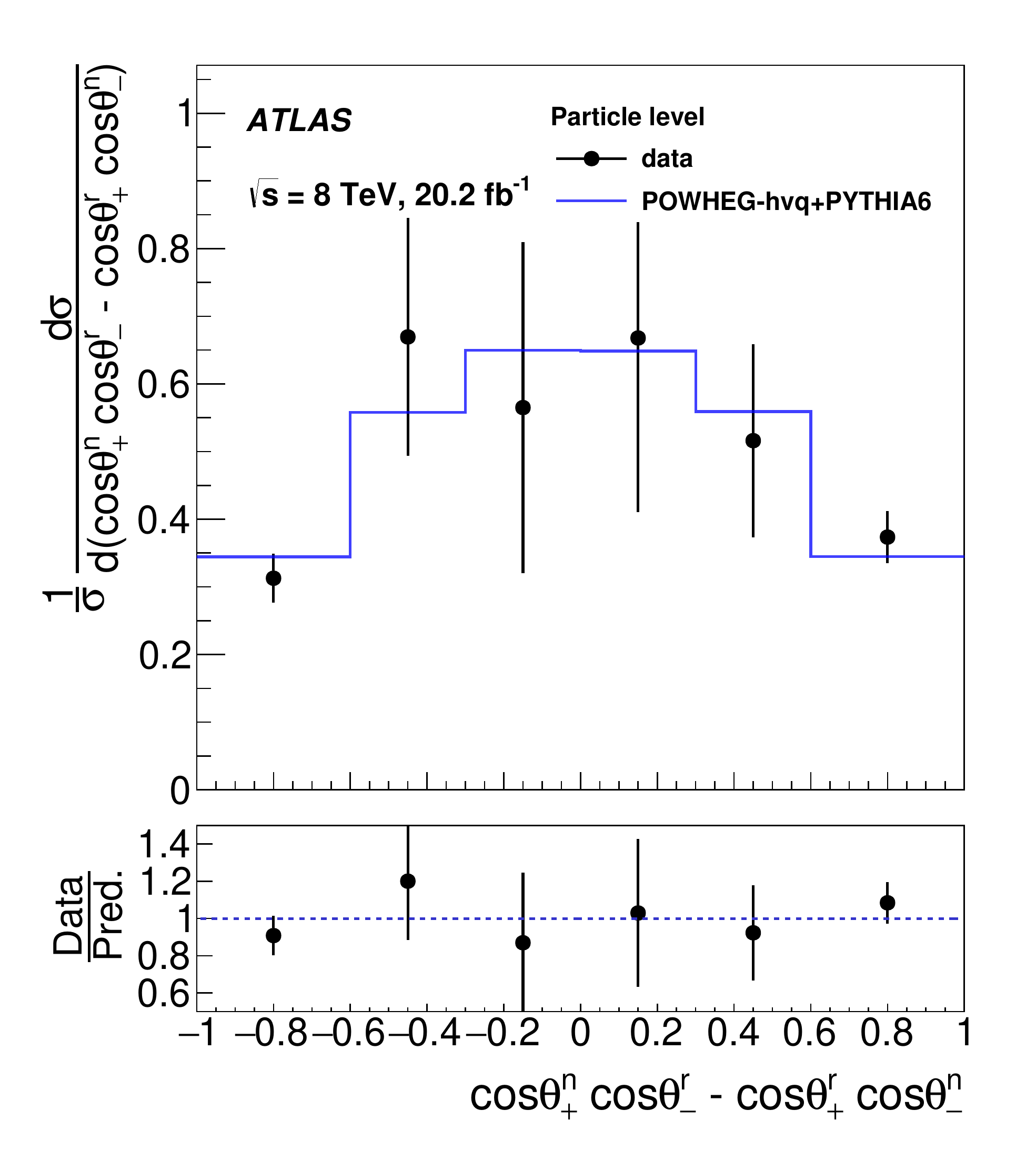} \\
   \includegraphics[scale=0.325]{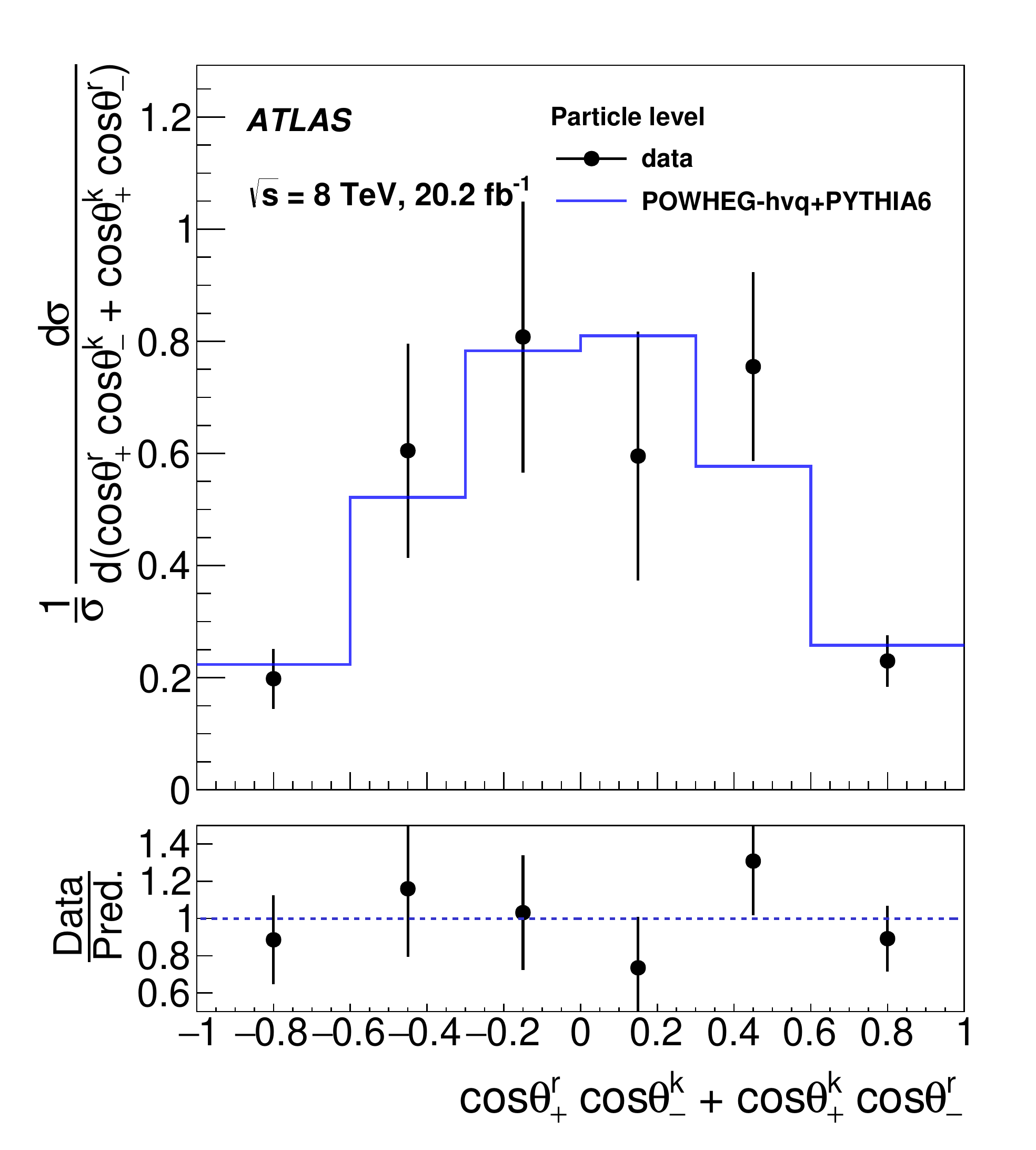}
   \includegraphics[scale=0.325]{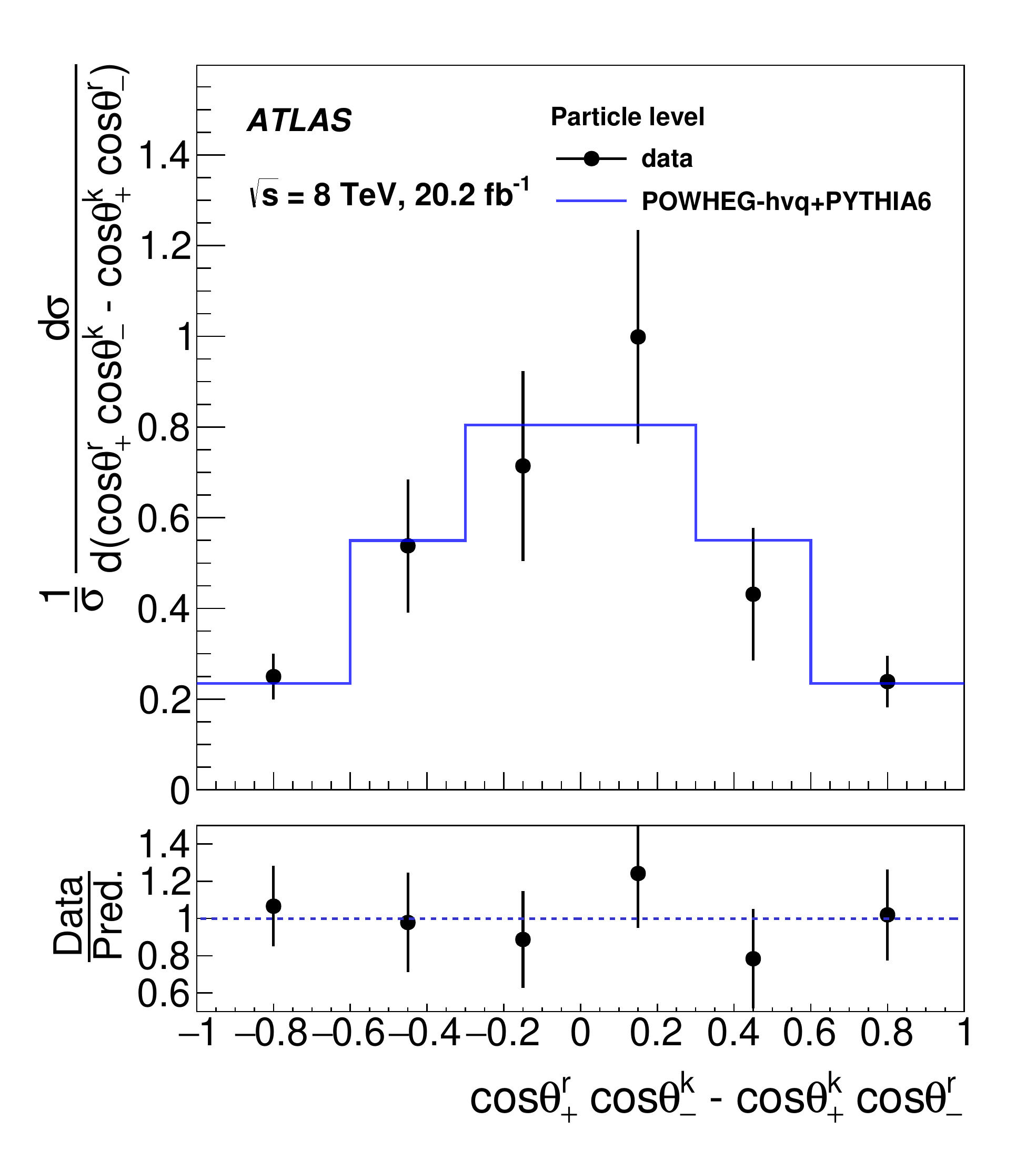} 
   \caption{Comparison of the unfolded cross correlation distributions and
     the prediction from the signal MC simulation for the stable-particle
     measurement. The total uncertainty is shown in each bin. 
     The bin-to-bin correlations between adjacent bins are typically between $-0.9$ and $-0.7$. 
     The correlations between non-adjacent bins range from $-0.4$ to 0.6.} 
   \label{fig:results_particle_crosscorr}
 \end{center}
\end{figure}

Figures~\ref{fig:results_particle_pol} to~\ref{fig:results_particle_crosscorr} show the observable distributions corrected
back to stable-particle level and compared to the generated
distribution created from \powheg\!+\pythia. 
No significant difference between the shapes of the observed and predicted distributions is observed.
The means of the distributions are compared between
unfolded data and MC predictions.
They are presented in Table~\ref{tab:results_summary}. 
In order
to compare the size of the uncertainties with the parton level measurement, the means
of the polarisation observables are multiplied by a factor of 3
and the correlations by a factor of $-9$ (Section~\ref{sec:observables}). 
Overall the total uncertainties for the measurements at parton and particle level are comparable.
The mass uncertainty is shown separately and not added to the total uncertainty, as explained in Section~\ref{sec:systematics}.
The dependence of the measured polarisations and spin correlations on the MC top quark
mass is presented in Table~\ref{tab:massSlope}. The measurements presented in this paper are compatible with other direct measurements in terms of
central values and uncertainties for the
polarisations along the helicity and transverse axis as well as for the
spin correlation along the helicity axis (Table~\ref{tab:measurements}).

\begin{center}
\begin{table}
\renewcommand\arraystretch{1.4}
\centering
\scriptsize
\begin{tabular}{|c|c|ccccc|c|}
\hline
Measurements & \text{Central}  & Total & Statistical & Detector & Modelling & Others & Mass\\ \hline 
\multicolumn{8}{c}{\rule{0pt}{4ex} Full phase-space}\\[3ex] \hline

$ \helpolplus $    & $-0.044$           &   $\pm 0.038$ &  $\pm 0.018$ &   $\pm 0.001$ &  $\pm 0.026$ &  $\pm 0.007$ &  $ \pm 0.027 $\\ 
$ \helpolminus $   & $-0.064$           &   $\pm 0.040$ &  $\pm 0.020$ &   $\pm 0.001$ &  $\pm 0.023$ &  $\pm 0.014$ &  $ \pm 0.027 $\\ 
$ \transpolplus $  & $-0.018$           &   $\pm 0.034$ &  $\pm 0.020$ &   $\pm 0.001$ &  $\pm 0.024$ &  $\pm 0.005$ &  -            \\ 
$ \transpolminus $ & $\phantom{+}0.023$ &   $\pm 0.042$ &  $\pm 0.020$ &   $\pm 0.001$ &  $\pm 0.034$ &  $\pm 0.005$ &  -            \\ 
$ \rpolplus $      & $\phantom{+}0.039$ &   $\pm 0.042$ &  $\pm 0.026$ &   $\pm 0.001$ &  $\pm 0.029$ &  $\pm 0.005$ &  -            \\ 
$ \rpolminus $     & $\phantom{+}0.033$ &   $\pm 0.054$ &  $\pm 0.023$ &   $\pm 0.002$ &  $\pm 0.045$ &  $\pm 0.006$ &  $ \pm 0.016 $\\ 
$ \helcorr $       & $\phantom{+}0.296$ &   $\pm 0.093$ &  $\pm 0.052$ &   $\pm 0.006$ &  $\pm 0.057$ &  $\pm 0.011$ &  $ \pm 0.037 $\\ 
$ \transcorr $     & $\phantom{+}0.304$ &   $\pm 0.060$ &  $\pm 0.028$ &   $\pm 0.001$ &  $\pm 0.047$ &  $\pm 0.001$ &  $ \pm 0.010 $\\ 
$ \rcorr $         & $\phantom{+}0.086$ &   $\pm 0.144$ &  $\pm 0.055$ &   $\pm 0.005$ &  $\pm 0.122$ &  $\pm 0.016$ &  $ \pm 0.038 $\\ 
$ \transhelsum $   & $-0.012$           &   $\pm 0.128$ &  $\pm 0.072$ &   $\pm0.005$ &  $\pm 0.087$ &  $\pm 0.029$ &  -             \\ 
$ \transheldiff $  & $-0.040$           &   $\pm 0.087$ &  $\pm 0.053$ &   $\pm0.004$ &  $\pm 0.058$ &  $\pm 0.003$ &  -             \\ 
$ \transrsum $     & $\phantom{+}0.117$ &   $\pm 0.132$ &  $\pm 0.070$ &   $\pm 0.003$ &  $\pm 0.102$ &  $\pm 0.010$ &  $ \pm 0.010 $\\ 
$ \transrdiff $    & $-0.006$           &   $\pm 0.108$ &  $\pm 0.069$ &   $\pm 0.005$ &  $\pm 0.070$ &  $\pm 0.004$ &  $ \pm 0.043 $\\ 
$ \rhelsum $       & $-0.261$           &   $\pm 0.176$ &  $\pm 0.083$ & $\pm0.006$ &  $\pm 0.135$ &  $\pm 0.011$ &  $ \pm 0.065 $   \\ 
$ \rheldiff $      & $\phantom{+}0.073$ &   $\pm 0.192$ &  $\pm 0.087$ &   $\pm 0.007$ &  $\pm 0.148$ &  $\pm 0.005$ &  $ \pm 0.025 $ \\ \hline

\multicolumn{8}{c}{\rule{0pt}{4ex} Fiducial phase-space}\\[3ex] \hline

 3$\langle \cos\theta_{+}^{k} \rangle$                                                           & $\phantom{+}0.125$ & $\pm 0.044$ &  $\pm 0.018$ &  $\pm 0.007$ &  $\pm 0.025$ &  $\pm 0.020$ &  $ \pm 0.027 $\\ %\hline
 3$\langle \cos\theta_{-}^{k} \rangle$                                                           & $\phantom{+}0.119$ & $\pm 0.040$ &  $\pm 0.022$ &  $\pm 0.008$ &  $\pm 0.021$ &  $\pm 0.014$ &  $ \pm 0.027 $\\ %\hline
 3$\langle \cos\theta_{+}^{n} \rangle$                                                           & $-0.025$           & $\pm 0.042$ &  $\pm 0.024$ &  $\pm 0.001$ &  $\pm 0.027$ &  $\pm 0.005$ &  -            \\ %\hline
 3$\langle \cos\theta_{-}^{n} \rangle$                                                           & $\phantom{+}0.023$ & $\pm 0.046$ &  $\pm 0.024$ &  $\pm 0.001$ &  $\pm 0.036$ &  $\pm 0.006$ &  -            \\ %\hline
 3$\langle \cos\theta_{+}^{r} \rangle$                                                           & $-0.104$           & $\pm 0.045$ &  $\pm 0.027$ &  $\pm 0.008$ &  $\pm 0.030$ &  $\pm 0.006$ &  -            \\ %\hline
 3$\langle \cos\theta_{-}^{r} \rangle$                                                           & $-0.110$           & $\pm 0.060$ &  $\pm 0.024$ &  $\pm 0.008$ &  $\pm 0.050$ &  $\pm 0.010$ &  $ \pm 0.015 $\\ %\hline
-9$\langle \cos\theta_{+}^{k}\cos\theta_{-}^{k} \rangle$                                         & $\phantom{+}0.172$ & $\pm 0.078$ &  $\pm 0.041$ &  $\pm 0.016$ &  $\pm 0.050$ &  $\pm 0.017$ &  $ \pm 0.027 $\\ %\hline
-9$\langle \cos\theta_{+}^{n}\cos\theta_{-}^{n} \rangle$                                         & $\phantom{+}0.427$ & $\pm 0.079$ &  $\pm 0.034$ &  $\pm 0.011$ &  $\pm 0.065$ &  $\pm 0.004$ &  $ \pm 0.027 $\\ %\hline
-9$\langle \cos\theta_{+}^{r}\cos\theta_{-}^{r} \rangle$                                         & $\phantom{+}0.031$ & $\pm 0.144$ &  $\pm 0.055$ &  $\pm 0.005$ &  $\pm 0.124$ &  $\pm 0.020$ &  $ \pm 0.033 $\\ %\hline
-9$\langle \cos\theta_{+}^{n}\cos\theta_{-}^{k} + \cos\theta_{+}^{k}\cos\theta_{-}^{n} \rangle$  & $\phantom{+}0.024$ & $\pm 0.132$ &  $\pm 0.078$ &  $\pm 0.004$ &  $\pm 0.085$ &  $\pm 0.025$ & -             \\ %\hline
-9$\langle \cos\theta_{+}^{n}\cos\theta_{-}^{k}-\cos\theta_{+}^{k}\cos\theta_{-}^{n} \rangle$    & $-0.047$           & $\pm 0.096$ &  $\pm 0.059$ &  $\pm 0.004$ &  $\pm 0.065$ &  $\pm 0.002$ & -             \\ %\hline
-9$\langle \cos\theta_{+}^{n}\cos\theta_{-}^{r} + \cos \theta_{+}^{r}\cos\theta_{-}^{n} \rangle$ & $\phantom{+}0.113$ & $\pm 0.143$ &  $\pm 0.076$ &  $\pm 0.005$ &  $\pm 0.108$ &  $\pm 0.023$ &  $ \pm 0.015 $\\ %\hline
-9$\langle \cos\theta_{+}^{n}\cos\theta_{-}^{r} - \cos \theta_{+}^{r}\cos\theta_{-}^{n} \rangle$ & $-0.030$           & $\pm 0.118$ &  $\pm 0.076$ &  $\pm 0.005$ &  $\pm 0.077$ &  $\pm 0.007$ &  $ \pm 0.052 $\\ %\hline
-9$\langle \cos\theta_{+}^{r}\cos\theta_{-}^{k} + \cos\theta_{+}^{k}\cos\theta_{-}^{r} \rangle$  & $-0.187$           & $\pm 0.151$ &  $\pm 0.069$ &  $\pm 0.023$ &  $\pm 0.122$ &  $\pm 0.006$ & $ \pm 0.039 $ \\ %\hline
-9$\langle \cos\theta_{+}^{r}\cos\theta_{-}^{k} - \cos\theta_{+}^{k}\cos\theta_{-}^{r} \rangle$  & $\phantom{+}0.047$ & $\pm 0.128$ &  $\pm 0.070$ &  $\pm 0.003$ &  $\pm 0.082$ &  $\pm 0.010$ & $ \pm 0.023$ \\ \hline

\end{tabular}
\caption{Results corrected to parton level in the full phase-space and to stable-particle level in the fiducial phase-space. 
  The central value with the total uncertainty is shown as well as the contribution from the various systematic uncertainty categories.
  The uncertainty from the "Background" category is not shown because it is always smaller than 0.001.
  The total uncertainty corresponds to the sum in quadrature of the uncertainty
  obtained from the unfolding procedure with marginalisation (including the
  background and detector modelling), the signal modelling and
  the "Others" category.
  The numbers shown for the "Detector" category correspond
  to the sum in quadrature of the individual estimates obtained as described in
  Section~\ref{sec:systematics}.
  The sum in quadrature of the values in the various columns thus does not necessarily match with the total uncertainty. 
  The uncertainty related to the top quark mass is presented separately. It is shown as "-" when found to be compatible with zero.
}
\label{tab:results_summary}
\end{table}
\end{center}

\begin{center}
	\begin{table}
		\centering
      \renewcommand{\arraystretch}{1.1}
		\begin{tabular}{ccc}
			\hline
			Measurements & Fiducial phase-space  & Full phase-space \\ \hline
			\helpolplus       & $ -0.04 \pm 0.01 $          & $ -0.04 \pm 0.01 $ \\ 
			\helpolminus      & $ -0.04 \pm 0.01 $          & $ -0.04 \pm 0.01 $ \\ 
			\transpolplus     & -                           & -                  \\ 
			\transpolminus    & -                           & -                  \\ 
			\rpolplus         & -                           & -                  \\ 
			\rpolminus        & $\phantom{+}0.02 \pm 0.01 $ & $ \phantom{+} 0.03 \pm 0.01 $ \\ 
			\helcorr          & $\phantom{+}0.04 \pm 0.01 $ & $ \phantom{+} 0.06 \pm 0.02 $ \\ 
			\transcorr        & $ -0.04 \pm 0.03 $          & $ -0.02 \pm 0.03 $ \\ 
			\rcorr            & $\phantom{+}0.05 \pm 0.03 $ & $ \phantom{+} 0.06 \pm 0.03 $ \\ 
			\transhelsum      & - & - \\ 
			\transheldiff     & - & - \\ 
			\transrsum        & $\phantom{+}0.02 \pm 0.01 $ & $ \phantom{+} 0.02 \pm 0.02 $ \\
			\transrdiff       & $ -0.08 \pm 0.01 $          & $ -0.07 \pm 0.01 $ \\ 
			\rhelsum          & $ -0.06 \pm 0.02 $          & $ -0.10 \pm 0.02 $ \\ 
			\rheldiff         & $\phantom{+}0.04 \pm 0.03 $ & $ \phantom{+} 0.04 \pm 0.03 $ \\ \hline
		\end{tabular}
    \caption{Dependence of polarisation and spin correlation measurements on the MC top quark mass. 
      The slope, computed from the reference value of 172.5 \GeV, is
      indicated for each measurement with its statistical uncertainty
      in units of $\GeV^{-1}$. %$\frac{1}{\text{\GeV}}$.
      Slopes which are compatible with zero within the uncertainty are indicated with "-".}
    \label{tab:massSlope}
	\end{table}
\end{center}

\begin{table}
  \scriptsize
  \begin{center}
    \centering
      \renewcommand{\arraystretch}{1.1}
    \begin{tabular}{lclccccc}
    \hline
      Experiment                  & $\sqrt{s}$ & Method       & \helpolplus        & \helpolminus        & \helcorr & \transpolplus & \transpolminus \\ \hline
      ATLAS                       & 8 \TeV     & Unfolding    & $-0.044 \pm 0.038$ & $-0.064 \pm 0.040$  & 0.296 $\pm$ 0.093 & $-0.018 \pm 0.034$ & 0.023 $\pm$ 0.042  \\ 
      CMS~\cite{spinpol_cms8TeV}  & 8 \TeV     & Unfolding    &  \multicolumn{2}{c}{$-0.022 \pm 0.058$}  & 0.278 $\pm$ 0.084 & -                  & -  \\ 
      ATLAS~\cite{pol_atlas}      & 7 \TeV     & Template fit &  \multicolumn{2}{c}{$-0.035 \pm 0.040$}  & -                 & -                  & - \\ 
      ATLAS~\cite{spin_atlas1}    & 7 \TeV     & Template fit & -                  & -                   & 0.23 $\pm$ 0.09   & -                  & - \\
      ATLAS~\cite{Aad:2015bfa}    & 7 \TeV     & Unfolding    & -                  & -                   & 0.315 $\pm$ 0.078 & -                  & - \\
      D0~\cite{pol_d0_ljets}      & 1.96 \TeV  & Template fit &  \multicolumn{2}{c}{ $-0.102 \pm 0.061$} & -                 & \multicolumn{2}{c}{0.040 $\pm$ 0.034} \\
     \hline
    \end{tabular}
    \caption{Direct measurements of polarisations or spin correlations for different experiments and measurement techniques. If more than one measurement from an experiment is performed with the same technique, the measurement with the smallest total uncertainty is shown. If a measurement quotes polarisation values for a CP-conserving and a CP-violating production mechanism, the result for the CP-conserving case is shown in the table (P and C denote the parity and charge-conjugation transformations, respectively). The template fits for the polarisation observables usually use the information of both the top and antitop quark decay chains. In this case, only one polarisation value can be quoted as the result and is shown for both columns of polarisation along the same axis. The SM predictions of the polarisations at the Tevatron are slightly different~\cite{Baumgart:2013yra,pol_d0_ljets} due to the different dominant production mechanism, which is \qqbar annihilation. Dashes indicate no measurement for the corresponding analysis.}
    \label{tab:measurements}
  \end{center}
\end{table}

% All figures and tables should appear before the summary and conclusion.
% The package placeins provides the macro \FloatBarrier to achieve this.
% \FloatBarrier
\clearpage

%-------------------------------------------------------------------------------
\section{Conclusion}
\label{sec:conclusion}
%-------------------------------------------------------------------------------

A measurement of 15 top quark spin observables was performed using a
data set of 20.2 $\text{fb}^{-1}$ of proton--proton collisions at
$\sqrt{s} = 8 $ \TeV, recorded by the ATLAS detector at the LHC.
The analysis is performed in the dilepton final state, characterised by the
presence of two isolated leptons.
Each of the observables is sensitive to a different coefficient of the spin
density matrix of \ttbar production.
The observable distributions are
corrected back to the parton and stable-particle level. At parton
level, the measurements along the helicity axis are
\begin{eqnarray}
\helpolplus & = & \ResultHelpolPlus \pm \TotalUncHelpolPlus~[\pm~\MassUncHelpolPlus~\textrm{(mass)} ],    \nonumber \\
\helpolminus & = & \ResultHelpolMinus \pm \TotalUncHelpolMinus~[\pm~\MassUncHelpolMinus~\textrm{(mass)} ], \nonumber \\
\helcorr & = & \phantom{+}\ResultHelcorr \pm \TotalUncHelcorr~[\pm~\MassUncHelcorr~\textrm{(mass)} ]. \nonumber
\end{eqnarray}
These values are in good agreement with the NLO SM predictions of
$\helpolplus = 0.0030 \pm 0.0010$, $\helpolminus = 0.0034 \pm 0.0010$ and $\helcorr = 0.318 \pm 0.003$.
The spin correlation along the transverse axis differs from zero
with a significance of 5.1 $\sigma$.
At stable-particle level, the unfolded distributions are compared to
their prediction from MC simulation (\powheg+\pythia). All distributions are
in agreement with the predictions.

%-------------------------------------------------------------------------------
\section*{Acknowledgements}
%-------------------------------------------------------------------------------

% Acknowledgements for papers with collision data
% Version 18-Jan-2017

% Standard acknowledgements start here
%----------------------------------------------
We thank CERN for the very successful operation of the LHC, as well as the
support staff from our institutions without whom ATLAS could not be
operated efficiently.

We acknowledge the support of ANPCyT, Argentina; YerPhI, Armenia; ARC, Australia; BMWFW and FWF, Austria; ANAS, Azerbaijan; SSTC, Belarus; CNPq and FAPESP, Brazil; NSERC, NRC and CFI, Canada; CERN; CONICYT, Chile; CAS, MOST and NSFC, China; COLCIENCIAS, Colombia; MSMT CR, MPO CR and VSC CR, Czech Republic; DNRF and DNSRC, Denmark; IN2P3-CNRS, CEA-DSM/IRFU, France; GNSF, Georgia; BMBF, HGF, and MPG, Germany; GSRT, Greece; RGC, Hong Kong SAR, China; ISF, I-CORE and Benoziyo Center, Israel; INFN, Italy; MEXT and JSPS, Japan; CNRST, Morocco; FOM and NWO, Netherlands; RCN, Norway; MNiSW and NCN, Poland; FCT, Portugal; MNE/IFA, Romania; MES of Russia and NRC KI, Russian Federation; JINR; MESTD, Serbia; MSSR, Slovakia; ARRS and MIZ\v{S}, Slovenia; DST/NRF, South Africa; MINECO, Spain; SRC and Wallenberg Foundation, Sweden; SERI, SNSF and Cantons of Bern and Geneva, Switzerland; MOST, Taiwan; TAEK, Turkey; STFC, United Kingdom; DOE and NSF, United States of America. In addition, individual groups and members have received support from BCKDF, the Canada Council, CANARIE, CRC, Compute Canada, FQRNT, and the Ontario Innovation Trust, Canada; EPLANET, ERC, ERDF, FP7, Horizon 2020 and Marie Sk{\l}odowska-Curie Actions, European Union; Investissements d'Avenir Labex and Idex, ANR, R{\'e}gion Auvergne and Fondation Partager le Savoir, France; DFG and AvH Foundation, Germany; Herakleitos, Thales and Aristeia programmes co-financed by EU-ESF and the Greek NSRF; BSF, GIF and Minerva, Israel; BRF, Norway; CERCA Programme Generalitat de Catalunya, Generalitat Valenciana, Spain; the Royal Society and Leverhulme Trust, United Kingdom.

The crucial computing support from all WLCG partners is acknowledged gratefully, in particular from CERN, the ATLAS Tier-1 facilities at TRIUMF (Canada), NDGF (Denmark, Norway, Sweden), CC-IN2P3 (France), KIT/GridKA (Germany), INFN-CNAF (Italy), NL-T1 (Netherlands), PIC (Spain), ASGC (Taiwan), RAL (UK) and BNL (USA), the Tier-2 facilities worldwide and large non-WLCG resource providers. Major contributors of computing resources are listed in Ref.~\cite{ATL-GEN-PUB-2016-002}.
%----------------------------------------------

%-------------------------------------------------------------------------------
\clearpage

%-------------------------------------------------------------------------------
\printbibliography
%-------------------------------------------------------------------------------
\clearpage

\newpage 
% ATLAS Collaboration author list
% Data extracted on 07-Feb-2017 for paper reference TOPQ-2015-13
%\documentclass[11pt]{article}
%\usepackage{a4wide}\begin{document}
\begin{flushleft}
{\Large The ATLAS Collaboration}

\bigskip

M.~Aaboud$^\textrm{\scriptsize 137d}$,
G.~Aad$^\textrm{\scriptsize 88}$,
B.~Abbott$^\textrm{\scriptsize 115}$,
J.~Abdallah$^\textrm{\scriptsize 8}$,
O.~Abdinov$^\textrm{\scriptsize 12}$,
B.~Abeloos$^\textrm{\scriptsize 119}$,
O.S.~AbouZeid$^\textrm{\scriptsize 139}$,
N.L.~Abraham$^\textrm{\scriptsize 151}$,
H.~Abramowicz$^\textrm{\scriptsize 155}$,
H.~Abreu$^\textrm{\scriptsize 154}$
% ...
\bigskip
\\
% ...
$^{8}$ Department of Physics, The University of Texas at Arlington, Arlington TX, United States of America\\
% ...
$^{12}$ Institute of Physics, Azerbaijan Academy of Sciences, Baku, Azerbaijan\\
% ...
$^{88}$ CPPM, Aix-Marseille Universit{\'e} and CNRS/IN2P3, Marseille, France\\
% ...
$^{115}$ Homer L. Dodge Department of Physics and Astronomy, University of Oklahoma, Norman OK, United States of America\\
% ...
$^{119}$ LAL, Univ. Paris-Sud, CNRS/IN2P3, Universit{\'e} Paris-Saclay, Orsay, France\\
% ...
$^{137}$ $^{(a)}$ Facult{\'e} des Sciences Ain Chock, R{\'e}seau Universitaire de Physique des Hautes Energies - Universit{\'e} Hassan II, Casablanca; $^{(b)}$ Centre National de l'Energie des Sciences Techniques Nucleaires, Rabat; $^{(c)}$ Facult{\'e} des Sciences Semlalia, Universit{\'e} Cadi Ayyad, LPHEA-Marrakech; $^{(d)}$ Facult{\'e} des Sciences, Universit{\'e} Mohamed Premier and LPTPM, Oujda; $^{(e)}$ Facult{\'e} des sciences, Universit{\'e} Mohammed V, Rabat, Morocco\\
% ...
$^{139}$ Santa Cruz Institute for Particle Physics, University of California Santa Cruz, Santa Cruz CA, United States of America\\
% ...
$^{151}$ Department of Physics and Astronomy, University of Sussex, Brighton, United Kingdom\\
% ...
$^{154}$ Department of Physics, Technion: Israel Institute of Technology, Haifa, Israel\\
$^{155}$ Raymond and Beverly Sackler School of Physics and Astronomy, Tel Aviv University, Tel Aviv, Israel\\
% ...
$^{a}$ Also at Department of Physics, King's College London, London, United Kingdom\\
$^{b}$ Also at Institute of Physics, Azerbaijan Academy of Sciences, Baku, Azerbaijan\\
$^{c}$ Also at Novosibirsk State University, Novosibirsk, Russia\\
$^{d}$ Also at TRIUMF, Vancouver BC, Canada\\
$^{e}$ Also at Department of Physics {\&} Astronomy, University of Louisville, Louisville, KY, United States of America\\
% ...
\end{flushleft}

%\end{document}
% Created with xml2latex.py

\end{document}